\documentclass[a4paper,fleqn,usenatbib]{mnras}

\usepackage{newtxtext,newtxmath}
\usepackage[T1]{fontenc}
\usepackage{ae,aecompl}

\usepackage[dvipsnames]{xcolor}

\usepackage{graphicx}	
\usepackage{amsmath}	
\usepackage{BibDef}
\usepackage{ragged2e}
\usepackage{booktabs}
\usepackage{multirow}
\usepackage{mathrsfs}
\usepackage{comment}
\usepackage{ulem}

\newcommand{\msun}{\ensuremath{\, \mathrm M_{\sun{}}}}

\newcommand{\ud}{~{\rm d}}

\title[The role of bars on the DF-driven inspiral of MOs]{The role of bars on the dynamical-friction driven inspiral of massive objects}

\author[E. Bortolas et al.]{%
Elisa Bortolas$,^{1,2,3}$\thanks{E-mail: elisa.bortolas@unimib.it }
Matteo Bonetti,$^{1,3,4}$
Massimo Dotti,$^{1,3,4}$
Alessandro Lupi,$^{1,5}$
\newauthor
Pedro R. Capelo,$^{2}$
Lucio Mayer$^{2}$ 
and
Alberto Sesana$^{1}$
\\
$^{1}$Dipartimento di Fisica ``G. Occhialini'', Università degli Studi di Milano-Bicocca, Piazza della Scienza 3, I-20126 Milano, Italy\\
$^{2}$Center for Theoretical Astrophysics and Cosmology, Institute for Computational Science, University of Zurich, Winterthurerstrasse\\ \ 190, CH-8057 Z\"urich, Switzerland\\
$^3$INFN, Sezione di Milano-Bicocca, Piazza della Scienza 3, I-20126 Milano, Italy\\
$^4$INAF, Osservatorio Astronomico di Brera, Via E. Bianchi 46, I-23807, Merate, Italy\\
$^5$Scuola Normale Superiore, Piazza dei Cavalieri 7, I-56126 Pisa, Italy\\
}

\date{Accepted XXX. Received YYY; in original form ZZZ}

\pubyear{2022}

\begin{document}
\label{firstpage}
\pagerange{\pageref{firstpage}--\pageref{lastpage}}
\maketitle

\begin{abstract}
In this paper, we {systematically} explore the impact of a galactic bar on the inspiral time-scale of a massive object (MO) within a Milky Way-like galaxy. We integrate the orbit of MOs in a multi-component galaxy model via a semi-analytical approach {that accounts } for dynamical friction  generalized to rotationally supported backgrounds. We compare the MO evolution in a galaxy featuring a Milky Way-like rotating bar to the evolution within an analogous axisymmetric galaxy without the bar. {In agreement with previous studies, we} find that the bar presence may significantly affect the inspiral, sometimes making it shorter by a factor of a few, sometimes hindering it for a Hubble time{. The erratic behaviour is mainly impacted by the relative phase at which the MO encounters the stronger bar-induced resonances. In particular,}
the effect of the bar is more prominent for initially in-plane, prograde MOs, especially those crossing the bar co-rotation radius or outer Lindblad resonance.  In the barred galaxy, we find the sinking of the most massive MOs ($\gtrsim 10^{7.5}\msun$) approaching the galaxy from large separations ($\gtrsim 8$ kpc) to be most efficiently hampered. Neglecting the effect of global torques associated with the non-symmetric mass distribution is thus not advisable even within {an idealized, smooth galaxy model; we further note that spiral patterns are unlikely to
affect the inspiral due to their transient and fluctuating nature.}
{We speculate that  the sinking  efficiency of massive black holes involved  in minor galaxy mergers may be hampered in barred galaxies, making them less likely to host a gravitational wave signal accessible to low-frequency detectors.
}
\end{abstract}



\section{Introduction}\label{sec:intro}
  
When a massive object (MO, an object with mass much larger than the typical  mass  of individual stars) orbits within its host galaxy, its trajectory is affected by the so-called dynamical friction \citep[DF, ][]{Chandrasekhar1942, Ostriker1999}. DF  arises as a response of the environment to the passage of the perturbing mass, and typically results in the gradual inspiral of the MO. In spite of the crude assumptions over which it has been first derived \citep[][]{Chandrasekhar1942,Chandrasekhar1943}, DF theory seems to properly describe the decay of many MOs, as galaxy satellites, stellar clusters, and massive black holes (MBHs) within their host systems \citep[e.g.][]{Inoue2009, Pfister2017}. However, most studies  supporting the success of DF theory model the host environment with  very simplistic and  idealized assumptions: the host galaxies are typically modelled as spherical and isotropic  or axisymmetric systems, with smooth  galactic potentials \citep[e.g.][]{Just2011, Arca-Sedda2014, Petts2015, Petts2016}. This is not surprising, as these are the systems that are typically addressed in standard (non cosmological) astrophysical simulations \citep[e.g.][]{White1978, Bortolas2016, Gualandris2017, Capelo_Dotti_2017, Bortolas2018, Bortolas2018tr, Tamfal2018}.   
Perhaps owing to this, DF alone has been often referred to as the very main phenomenon capable to determine the decay of an orbiting MO \citep[e.g.][]{Tremaine1975, Begelman1980}. 

Only recently, a number of studies started exploring the evolution of MBHs within  much less idealized galactic environments, featuring, e.g. the cosmological evolution of the galaxies, the possible formation of clumps, spirals, bars, the effect of star formation and hydrodynamics and so forth \citep[e.g.][]{Fiacconi2013, VanWassenhove_et_al_2014, Lupi2015, Roskar2015, Tamburello2017,Souza-Lima2017, Tremmel2018, Tremmel2018b, Pfister2019, Bellovary_et_al_2019, Bortolas2020, Souza-Lima2020}. The MO evolution within these more realistic, composite galaxies appears to be much harder to predict in the DF framework, as the aforementioned non-symmetric, time-dependent perturbations in the potential result in a much more erratic orbital evolution. 

\citet{Bortolas2020} evolved a set of MBHs within a typical, irregular and turbulent galaxy at $z\gtrsim 6$, embedded in a cosmological environment. They found that, once a strong bar develops in the host galaxy, the MBHs orbital evolution is critically affected by it: owing to the bar interaction, the decay time is $\sim10$ times faster than what DF theory would predict for four out of five MBHs, while in one case the interaction  kicks the MBH in the galaxy outskirts \citep[][]{Bortolas2020}. This study further highlights that the magnitude of the galactic global torques resulting from the non-symmetric galactic mass distribution is virtually always much stronger than the DF-induced torques, suggesting that assuming DF to be the main driver of the inspiral may be inadequate  for realistic galaxies \citep[][]{Bortolas2020}.

The aforementioned shortcomings of DF theory are particularly relevant in view of the  forthcoming opening of a low-frequency ($<0.1$ Hz) gravitational wave window,  where the nano-Hz regime is being probed by Pulsar Timing Arrays (PTAs; \citealt{2016MNRAS.458.3341D,2016MNRAS.455.1751R,2019MNRAS.490.4666P,2021ApJS..252....5A}), and the milli-Hz band will be explored by the Laser Interferometer Space Antenna (LISA; \citealt{Amaro-Seoane2017,Schodel2017LISA,Barack_et_al_2019}) in the 2030s.   Therefore, it is important to constrain the  time spanning from a galaxy merger to the gravitational wave induced coalescence of the host's MBHs, that is going to be observed by the aforementioned facilities; such time-scales would obviously strongly depend on the physics of the large scale galactic inspiral.

In this paper, we aim at addressing more systematically how galactic bars affect the decay time-scale of MOs. Conservatively, we explore the evolution of MOs in an idealized, Milky Way-like galaxy, in which the only deviation from axisymmetry is constituted by a rotating triaxial bar of $\approx5$ kpc extension.
We integrate the MO orbit with the semi-analytical code presented in \citet{Bonetti2020, Bonetti2021}, whose novel treatment for DF  guarantees remarkable agreement with $N$-body simulations of composite galaxies. 
We perform a large number of numerical experiments, comparing the decay time-scale in the barred galaxy to its value in an analogous, axisymmetric galaxy not featuring the bar. 
In Sec.~\ref{sec:methods}, we detail the methodology adopted for the orbits initialization and integration; Sec.~\ref{sec:theory} briefly reviews the theoretical aspects related to the orbital evolution within a uniformly rotating, non-axisymmetric potential; in Sec.~\ref{sec:results}, we present the results of our study, which are then discussed and summarized in Sec.~\ref{sec:concl}.


\section{Methods}\label{sec:methods}

\subsection{Galaxy potential}

\begin{table*}
  \centering
  \caption{Reference galaxy structural parameters}
  \label{tab:mwstruct}
  \begin{center}
      
    \begin{tabular}{ccccc}
        \hline
        Component & Model & Mass [$\msun$] & scale length [kpc] & others
        \\ \hline
        Bulge$^\ast$ & \citet{Dehnen1993} & $M_{\rm B} = 5\times 10^9$ & $r_{\rm B} = 0.7$ & $\gamma = 1$ \\%
        Disc$^\ast$ & Exponential \citep{BinneyTremaine2008} & $M_{\rm D} = 3\times 10^{10}$ & $(r_{\rm D}, z_{\rm D}) = (3, 0.3)$ & -- \\%
        Halo & \citet*{Navarro1996} & $M_{\rm H} = 4.317\times 10^{11}$, $M_{\rm V} = 8\times 10^{11}$ & $r_{\rm H} = 16$, $r_{\rm V} = 245$  & $c_{\rm H} = 15.3$ \\%
        Bar & \textit{Softened Needle} \citep{Long1992} & $M_{\rm bar} = 1.8\times 10^{10}$ & $(a,b,c)_{\rm bar} = (5,2,0.3)$ & $\omega_{\rm bar} = 40$ km s$^{-1}$ kpc$^{-1}$ \\%
         \hline
    \end{tabular}
    
  \end{center}
  \justifying
  {\footnotesize $^\ast$ Note that the disc and bulge mass shown here refer to the case in which the bar is present, and have to be enhanced as discussed in the text for the integrations that are not featuring a bar.}
\end{table*}

We first introduce the reference parameters adopted for the study of the Milky Way-like galaxy. We model the galaxy by considering components of different shape and nature, specifically: a stellar spherical bulge, a stellar disc, and (in part of our runs) a stellar bar, all of them embedded in a dark matter halo.
The properties for the different Galactic components adopted here are in agreement with recent literature on the topic, and in particular with \citet{Bovy2015}; the properties of the Galactic bar are taken from \citet{Portail2017}. Table~\ref{tab:mwstruct} reports the relevant values adopted for the galaxy initialization.

Specifically, we represent the central bulge using a \citet{Dehnen1993} potential well, whose associated density profile reads
\begin{equation}
    \rho_{\rm B}(r) = \dfrac{(3-\gamma)M_{\rm B}}{4\pi} \dfrac{r_{\rm B}}{r^\gamma (r+r_{\rm B})^{3-\gamma}},
    \label{eq:bulge}
\end{equation}
where $M_{\rm B}$ is the bulge total mass, $r_{\rm B}$ is its characteristic radius, and $\gamma$ represents the inner density slope of the model; finally, $r$ is the distance from the centre. We choose $\gamma=1$, which corresponds to a \citet{Hernquist1990} profile. 

The disc is modelled with an exponential profile \citep{Spitzer1942,BinneyTremaine2008}: 
\begin{equation}
  \rho_{\rm D}(R,z) = \dfrac{M_{\rm D}}{4\pi r_{\rm D}^2 z_{\rm D}} {\rm e}^{-R/r_{\rm D}} {\rm sech}^2\left(\dfrac{z}{z_{\rm D}}\right),  
  \label{eq:disc}
\end{equation}
where $R$ represents the cylindric radius, $z$ is the coordinate perpendicular to the disc, $r_{\rm D}$ is the disc scale lenght, $z_{\rm D}$ is the disc scale height, and $M_{\rm D}$ is the total mass of the disc. Given that an analytical expression for the associated disc potential does not exist, the integrator obtains the accelerations induced by the disc potential numerically, as described in detail in \citet{Bonetti2021}. In order to speed up the computation of the disc acceleration, the $R$ and $z$ components of the acceleration are tabulated in an adaptive grid spanning several orders of magnitude in both $R$ and $z$, and the acceleration along the integration is obtained by  interpolating the tabulated values for the $R$ and $z$ values needed at each timestep.

The dark matter halo is described via a \citet{Navarro1996} potential, whose associated density  profile is
\begin{equation}
    \rho_{\rm H}(r) = \dfrac{M_{\rm H}}{4 \pi r_{\rm H}^3} \dfrac{r_{\rm H}}{r (1+r/r_{\rm H})^2},
    \label{eq:halo}
\end{equation}
where $M_{\rm H}$ is the mass scale of the model and $r_{\rm H}$ its scale radius; the model virial mass can be expressed as $M_{\rm V}= M_{\rm H} [\ln(1+c_{\rm H})-c_{\rm H}/(1+c_{\rm H})]$, with $c_{\rm H}$ concentration parameter defined as the ratio between the galaxy virial radius $r_{\rm V}$ and $r_{\rm H}$.
The above three components (central bulge, disc, and halo) are always accounted for in our study. 

In addition to the disc, bulge, and halo components, in many of our runs we also account for the presence of a galactic bar. 
We model the bar as a \textit{softened needle} profile \citep{Long1992}, whose potential has the form
\begin{align}
        \Phi_{\rm bar}(x, y, z) &= \dfrac{GM_{\rm bar}}{2a_{\rm bar}}\ln{\left(\dfrac{x-a_{\rm bar}+T_-}{x+a_{\rm bar}+T_+}\right)}\\
        T_{\pm} &=\{(a_{\rm bar}\pm x)^2 + y^2 + [b_{\rm bar} + (c_{\rm bar}^2+z^2)^{1/2}]^2\}^{1/2},
        \label{eq:bar_potential}
\end{align}
where $G$ is the gravitational constant, $M_{\rm bar}$ is the total mass of the bar, and $(a_{\rm bar}, b_{\rm bar},c_{\rm bar})$ are the scale lengths in the direction of the $(x,y,z)$ Cartesian coordinates. We  assume the bar to initially  lie along the $x$ direction. The bar rotates in the disc plane with a constant orbital frequency $\omega_{\rm bar}$.
The parameters associated to the bar are also shown in Table~\ref{tab:mwstruct}.  Note that the mass of the disc and bulge  are adjusted depending on the presence (or absence) of the bar.

In order to disentangle the effect of the bar alone on the evolution of MOs, we decided to run all our integrations in two analogous galaxy models, one featuring the galaxy bar described above, and another one which is purely axisymmetric, and in which the mass assigned to the bar is re-distributed between the bulge and disc components. We perform this latter task by redistributing the bar mass%
  \footnote{Note that the prescription we propose for redistributing the bar mass into the disc and bulge is by no means a general prescription and we found it to work well for the galaxy we are considering, but it may fail if different galaxy potentials are adopted.}  
as: $M_{\rm B} \rightarrow  M_{\rm B}+0.1\times M_{\rm bar} (r_{\rm B}/a_{\rm bar})$,  $M_{\rm D} \rightarrow  M_{\rm D}+ M_{\rm bar}(1 - 0.1\times r_{\rm B}/a_{\rm bar})$.
We find that this choice allows us to maintain a very similar rotation curve in the disc plane for the two models: if the circular velocity of the barred galaxy in the disc plane is averaged over all possible bar orientations, we find our prescription to maintain the  deviation between the two always below 4 per cent (see Fig.~\ref{fig:rot_curve} in the Appendix). For clarity, in the following we will always refer to the galaxy rotation curve as that associated  to the barred galaxy.

The characteristic {resonances} of the galaxy are shown in Table~\ref{tab:resonances}. The profiles of the epicyclic frequency and the orbital frequency are computed in the non-barred galaxy, and the bar orbital frequency is used to define the various resonances reported in the Table.

\subsection{Dynamical friction prescriptions}

The implementation for the DF-induced deceleration suffered by the MO along its orbital evolution is described in detail in \citet{Bonetti2020, Bonetti2021}. Here we report the key aspects of the implementation, and we refer the reader to the aforementioned papers for more details.

The DF acting on the MO is computed as a sum of the DF associated to the different galactic components. Each of the spherically symmetric components induces a deceleration with the form
\begin{equation}\label{eq:DF_sph}
    \mathbf{a}_{\rm df,sph} = -2\pi G^2 \ln(1+\Lambda^2) m_{\rm p} \rho(r) \left({\rm erf}(X) - \dfrac{2 X{\rm e}^{-X^2}}{\sqrt{\pi}}\right) \dfrac{\mathbf{v}_{\rm p}}{|\mathbf{v}_{\rm p}|^3},
\end{equation}
where $m_{\rm p}$ and $\mathbf{v}_{\rm p}$ are the MO mass and instantaneous velocity, respectively, $\rho(r)$ is the local background density associated with the given spherical component, and $X=v_{\rm p}/(\sqrt{2}\sigma(r))$, with $\sigma(r)$ being the local velocity dispersion of the considered galactic component. The argument of the Coulomb logarithm in the equation is given by the ratio between the maximum and minimum impact parameters, $\Lambda = p_{\rm max}/p_{\rm min}$, computed as $p_{\rm max} = r\left(- \ud \ln \rho/\ud \ln r\right)^{-1}$ and $p_{\rm min} = \max[{G m_{\rm p}}/\left({v_{\rm p}^2+\sigma(r)^2}\right), D_{\rm p}]$, where $D_{\rm p}$ is the physical radius of the MO (which we set to zero in the present integration, as we always assume non extended MOs, as MBHs).

The DF associated to the rotating disc is modelled as
\begin{align}\label{eq:adf_disc}
    \mathbf{a}_{\rm df,disc} = -2\pi G^2 \ln(1+\Lambda^2) & m_{\rm p} \rho_{\rm D}(R,z) \ \times \nonumber\\
    & \times\left({\rm erf}(X_{\rm D}) - \dfrac{2 X_{\rm D}{\rm e}^{-X_{\rm D}^2}}{\sqrt{\pi}}\right) \dfrac{\mathbf{v}_{\rm rel}}{|\mathbf{v}_{\rm rel}|^3},
\end{align}
where $\mathbf{v}_{\rm rel} = \mathbf{v}_{\rm p} - \mathbf{v}_{\rm rot}(R)$,   and $\mathbf{v}_{\rm rot}(R)$ is the rotational velocity in the disc, generally not equal to the circular velocity of the disc, as we assume that it is not fully rotationally supported. Finally, $X_{\rm D} = v_{\rm rel}/(\sqrt{2}\sigma_{\rm R})$, where $\sigma_{\rm R}$ denotes the radial velocity dispersion of the disc.
The details for the computation of $\mathbf{v}_{\rm rot}(R)$ and $\sigma_{\rm R}$ can be found in \citet[][]{Bonetti2021}.
In the above expression, the minimum impact parameter entering the Coulomb logarithm is  computed as $ p_{\rm min, D} = G m_{\rm p}/(v_{\rm rel}^2 + \sigma_{\rm R}^2)$, whereas $p_{\rm max}$ is chosen equal to the disc scale height. Physically, the expression adopted for the DF in the rotating disc accounts for the fact that the MO is moving within a medium featuring a net rotational velocity, so that the relative velocity between the MO and the disc typical rotational velocity at each given radius is what has to be accounted for when estimating the MO deceleration. This treatment has been proven to work very well in rotating environments \citep[][]{Bonetti2021}, and it reproduces  the so-called drag-towards-circular-corotation (see Sec.~\ref{sec:counterrotating} for a description).

The DF induced by the bar, when present, is very difficult to describe starting from first principles. Here we considered only the effect produced by the enhanced density and the additional DF caused by the bar is simply obtained by adding the bar density to the disc  component in the equations for the deceleration (Eq.~\ref{eq:adf_disc}). Note that this assumption can in principle be inaccurate and impact our results. In Appendix~\ref{sec:appB}, we thus compare our semi-analytical treatment with full $N$-body simulations. The stochasticity induced by merely changing the number of particles in the $N$-body run is significant, thus suggesting that the detailed implementation of a more accurate DF prescription would probably not severely impact the evolution, as stochasticity induced by the fact that the orbits are chaotic appears to be  the main factor in determining the decay time-scale.


\subsection{Initial conditions for the orbit}

\begin{figure}
\centering
\includegraphics[ width=0.45\textwidth]{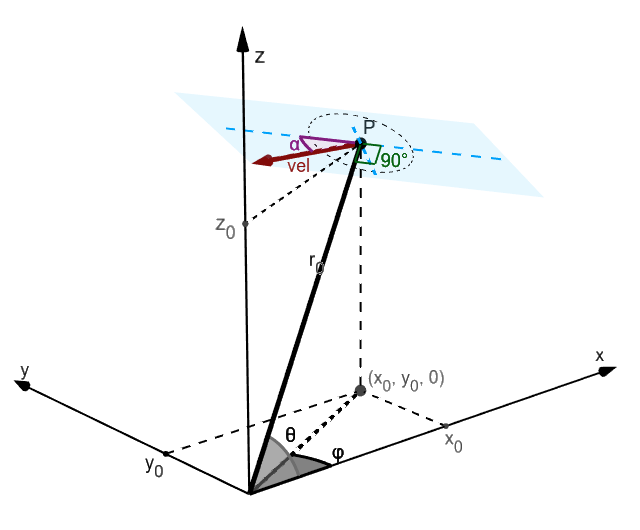}
    \caption{The image shows the relevant variables adopted to initialize the orbit of the MO in the presented integrations. The point P $(x_0, y_0, z_0$) in which the MO is initialized is defined by the azimuthal and polar angles $\theta$, $\phi$, and by the length  $r_0$ of the position vector. The velocity (${\rm vel}$, indicated as $v_{\rm p}$ in the text) always lies in the plane perpendicular to the position vector associated to P, and its orientation is defined by the angle $\alpha$, which is defined to be $0$ if the velocity lies parallel to the $x-y$ plane.  }
    \label{fig:orbit_initialization}
\end{figure}

In order to explore the effect of bars on the MOs dynamics, we perform a large number of orbital integrations. Each of the simulations   is always performed with the very same initial conditions in the galaxy featuring and non-featuring the galactic rotating bar. The MO does not suffer any mass variation during the evolution; this is obviously a simplification, and we plan to implement the effect of the MO mass loss in a forthcoming study.  The initialization of the orbit of the MO  is characterized by a series of variables that serve to uniquely determine the initial position and velocity of the MO, and specifically we will mainly use the following:
\begin{itemize}
    \item $r_0$, the initial distance of the MO from the centre of the system;
    \item $f_{\rm circ} \in (0,1]$: if $v_{\rm c}$ is the circular velocity at $r_0$, we assign to the MO a tangential velocity equal to $v = f_{\rm circ}v_{\rm c}$, and zero radial velocity, meaning that the orbital evolution in the axisymmetric case and in the disc plane is always initialized at the apocentre;\footnote{Note that the apocentre is not well defined out of the disc plane and in the barred case.} so  $f_{\rm circ} \simeq 0$ means an almost radial orbit, and $f_{\rm circ}=1$  corresponds to an ideally circular orbit (when the MO is in the disc plane).  Note that $v_{\rm c}$ is always taken in the disc plane, regardless of whether the MO actually starts its evolution within the disc;
    \item $\phi \in [0, 180)$ degrees, the azimuthal angle; in principle, this should run from 0 to 360 degrees, but we limit its range for symmetry reasons. Note that this angle can be neglected for the non-barred galaxy, as the potential is axisymmetric. In the barred case, $\phi=0$ means that the MO initially sits along the bar longest principal axis, $a_{\rm bar}$;
    \item $\theta \in [0, 180]$  degrees, the  angle between the disc  ($x-y$) plane and the MO initial position vector;
    \item $\alpha \in [0, 360)$  degrees, the angle between the initial velocity vector and the $x-y$ plane; remember that the initial velocity vector is always perpendicular to the position vector of the particle;
    \item $i \in [0, 180)$ degrees, the inclination of the initial orbit with respect to the disc plane. Note that this variable is degenerate with the previous three angular variables, but we will refer to it as well in some situations.
\end{itemize}
%
Fig.~\ref{fig:orbit_initialization} shows most of the aforementioned quantities in the three-dimensional space.
In what follows, we define the inspiral to be completed once the MO stably remains below a separation of $10$ pc from the centre;  we always stop the integration when the simulation time reaches a Hubble time (assumed to be 13.7 Gyr).

\begin{table}
  \centering
  \caption{Characteristic scales for resonances}
  \label{tab:resonances}
  \begin{center}
      
    \begin{tabular}{lc}
        \hline
        Label & Value 
        \\ \hline
        Bar semi-major axis &  5.0000 kpc \\%
        Co-rotation radius & 5.0041 kpc \\%
        Inner Lindblad resonance & 0.5746 kpc \\%
        Outer Lindblad resonance & 8.8870 kpc \\%
        Saddle radius ($\Phi_{\rm eff}$) & 5.3165 kpc \\%
        Maxima radius ($\Phi_{\rm eff}$) & 4.9620 kpc \\%
         \hline
    \end{tabular}
    
  \end{center}
  \justifying
  {\footnotesize The table displays the characteristic scales at which the bar resonates with the galaxy characteristic orbital frequencies, and the radii of the saddle points and maxima associated to the effective potential shown in Fig.~\ref{fig:bar_contour}.}
\end{table}


\section{Theoretical background}\label{sec:theory}

\begin{figure}
\centering
\includegraphics[ width=0.45\textwidth]{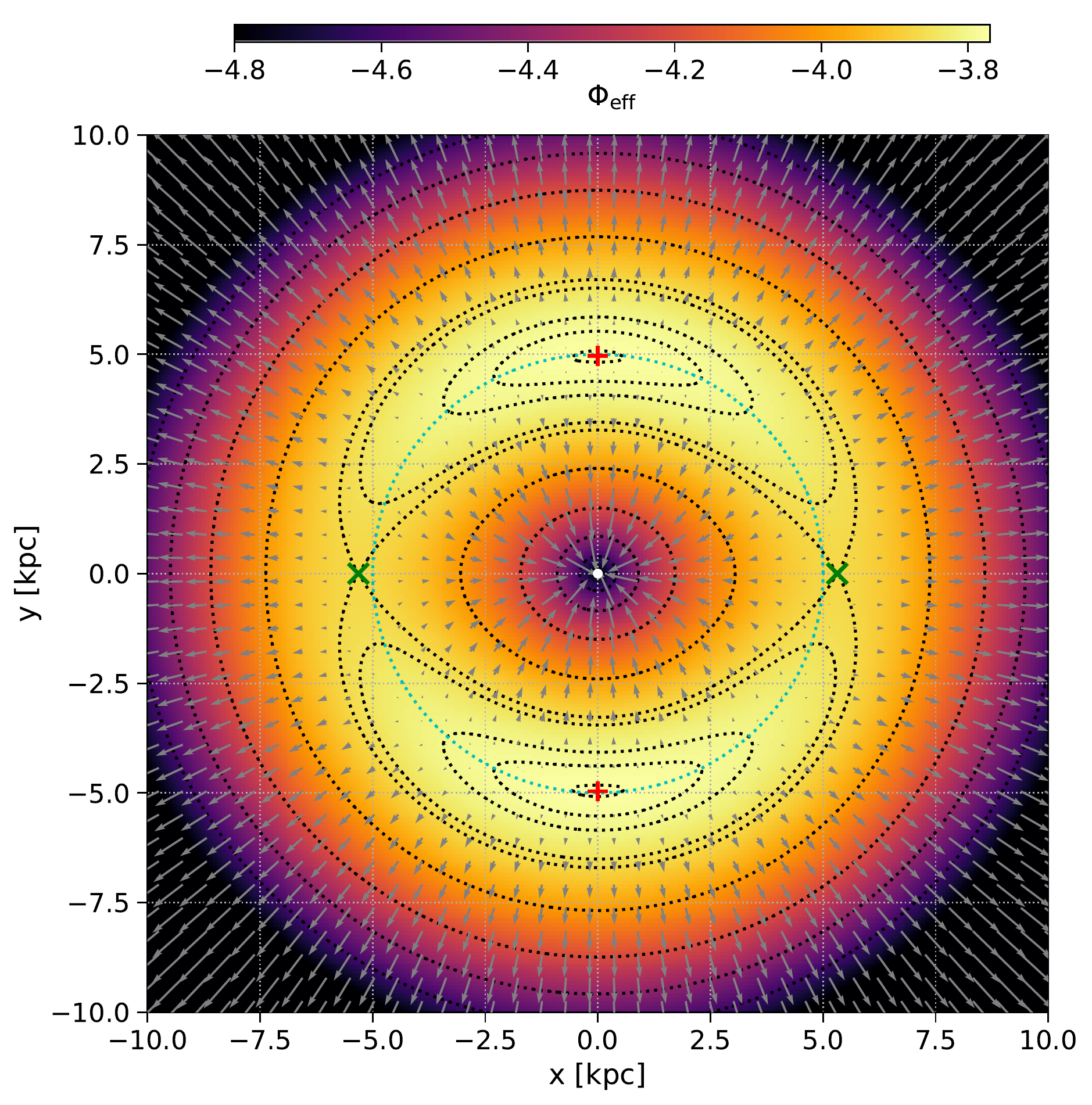}
    \caption{The colour-coded map displays the effective galactic potential $\Phi_{\rm eff}$ in the plane of the disc ($z=0$), measured in units of $4.301\times10^4 $ km$^2$ s$^{-2}$. The central white point in the origin marks a central minimum, the two red `+' are the two potential maxima ($x=0; y\approx\pm 4.962$~kpc), and the green `$\times$' are the two saddle points ($x\approx\pm 5.316$~kpc; $y=0$). The cyan line describes a circle of radius 5 kpc, i.e. the spatial extension of the bar; the arrows indicate the direction of the gravitational force associated to the effective potential displayed, with their length being proportional to its magnitude.}
    \label{fig:bar_contour}
\end{figure}

In order to understand the  behaviour of the MO evolution, we recall that the gravitational potential of a uniformly
rotating, non-axisymmetric density distribution is usefully described in a framework that co-rotates with the triaxial perturbation. In particular, it is useful to define the effective potential \citep[e.g.][]{Sellwood1993}
\begin{equation}
\Phi_{\rm eff} = \Phi-\frac 1 2 \omega_{\rm bar}^2 r^2,    
\end{equation}
where $\Phi$ is the conservative, space-dependent galactic potential of the barred galaxy, $\omega_{\rm bar}$ is the rotational frequency of the bar and $r$ is the distance from the centre. Even neglecting the Coriolis force (which depends on the velocity of the moving mass), the gradient of the effective potential gives a  good estimate of the force experienced by a test mass in the rotating frame within the disc plane. Fig.~\ref{fig:bar_contour} displays a map of the effective potential in the plane of the disc, for  our barred galaxy model; the plot additionally shows the magnitude and direction of the associated effective force. This effective potential can be thought as a `volcano' \citep{Prendergast1983}, with a minimum (crater) at the
centre, a rim whose height varies slightly, and the   slope that descends at larger radii. In this framework, neglecting DF, the so-called Jacobi integral (rather than the standard  energy) of a test mass in the disc plane is conserved in time. This quantity can be expressed as 
\begin{equation}\label{eq:EJ}
E_{\rm J} = E  - \omega_{\rm bar}J_z = v_p^2/2  + \Phi_{\rm eff},
\end{equation}
where $E$ and $J_z$ are, respectively, the energy and $z$ component of the angular momentum per unit mass, and $v_{\rm p}$ is the velocity magnitude, all quantities being measured  in the non-rotating frame; note that $E_J$ is defined in the plane of the disc. In absence of DF,  $E_{\rm J}$ would determine whether a mass is limited to orbits in a particular region of space: only  if $E_{\rm J}$ is larger than the maxima of $\Phi_{\rm eff}$, it can in principle explore the entire galaxy plane. 
It is also relevant to note that the two saddle points are unstable equilibria points, whereas in the present galaxy model the two potential maxima and the central minimum are stable points, meaning that a test mass can stably sit there or orbit these points  in the absence of perturbations. More details on the orbits of subject masses in triaxial, rotating potentials can be found in, e.g. \citet[][especially from their sec.~4.3.2]{Sellwood1993}.

In our present framework, the otherwise conserved $E_{\rm J}$, that determines the orbit of a subject mass, can vary  due to the effect of DF. The above considerations allow us to better understand the orbital behaviour of MOs subject to the combined effect of the galactic potential, the rotating bar, and  DF.

\section{Results}\label{sec:results}

\begin{figure*}
\centering
\includegraphics[ width=0.49\textwidth]{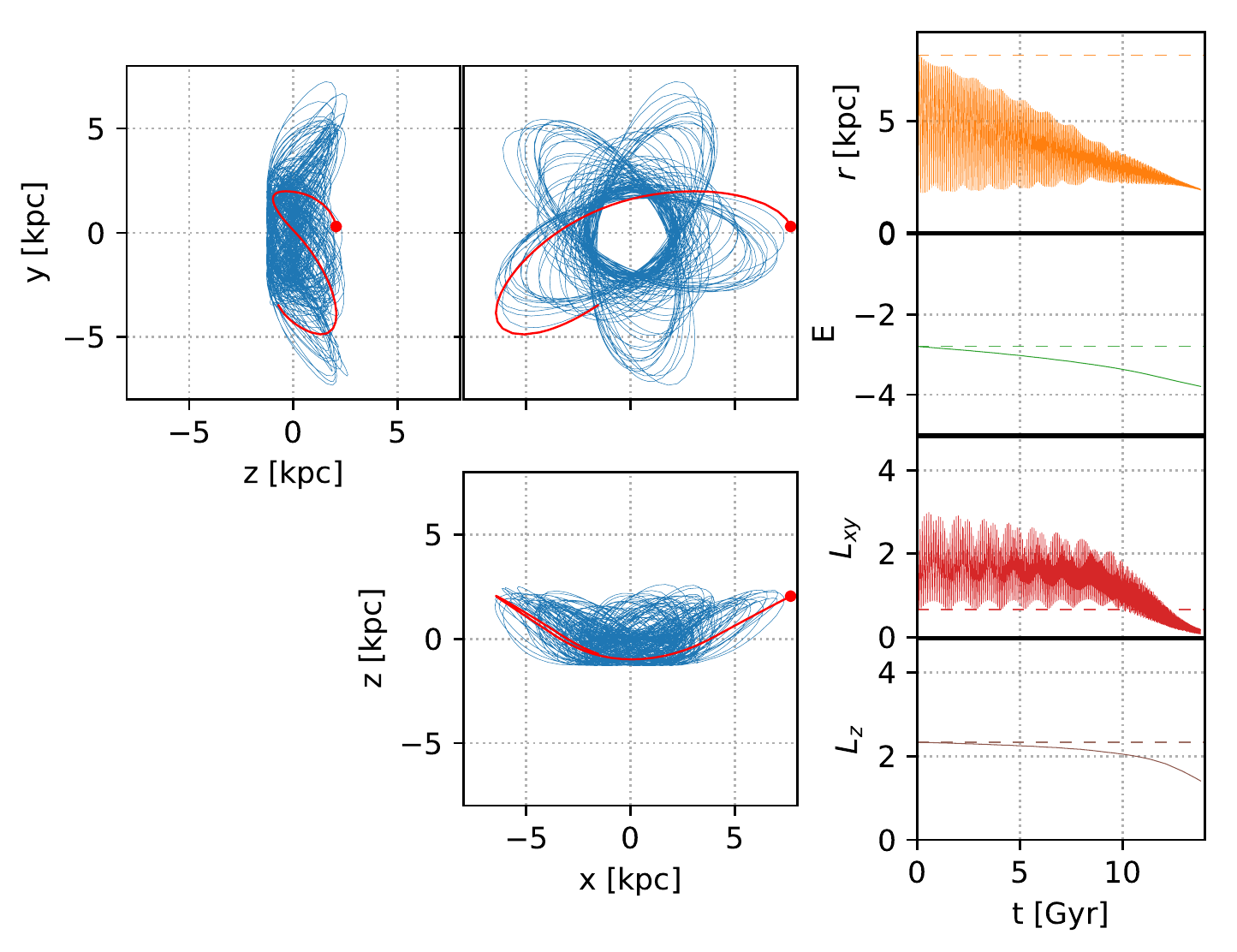}
\includegraphics[ width=0.49\textwidth]{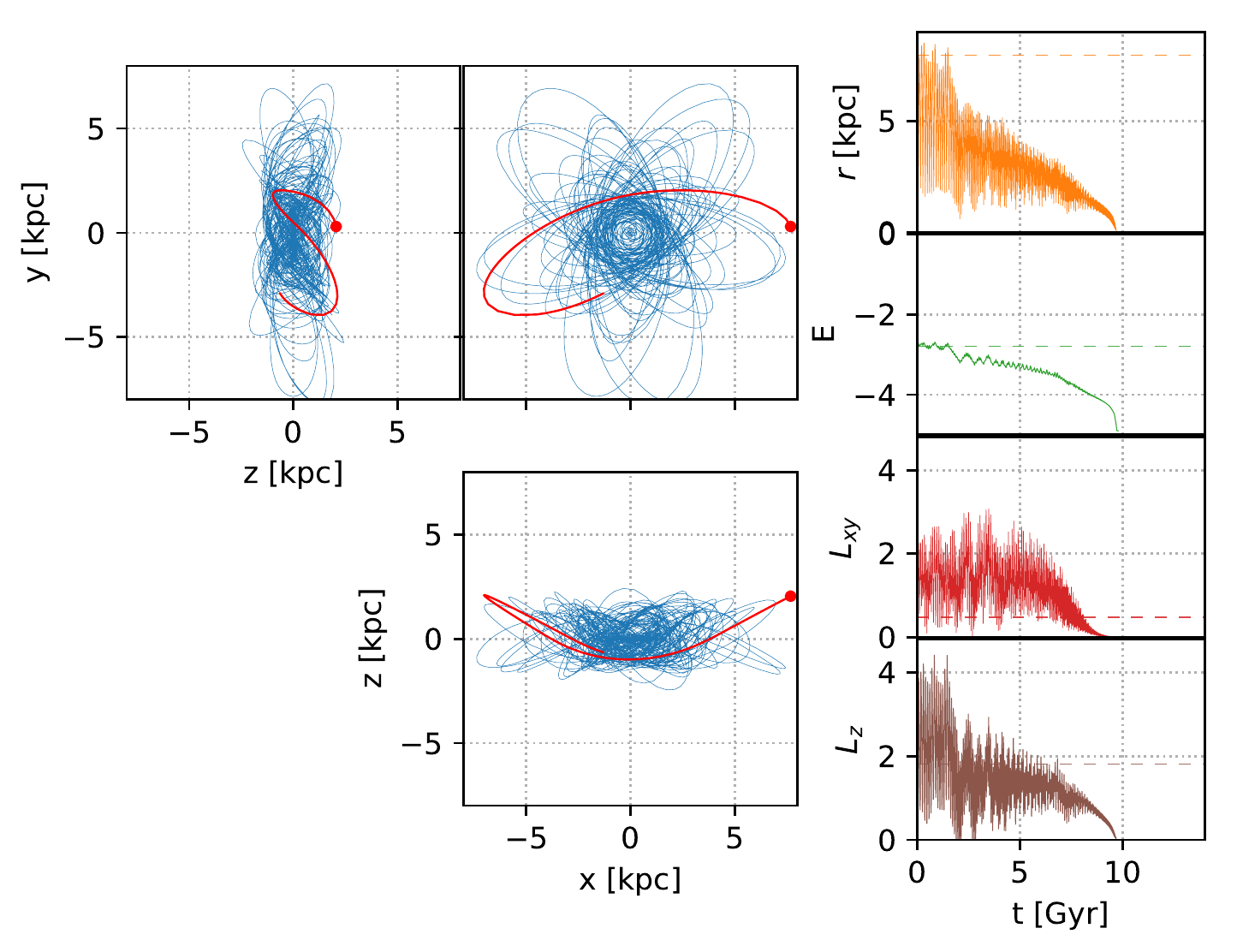}
    \caption{The plot shows various quantities associated to the orbital evolution of an MO  in the non-barred (left-hand panels) and barred (right-hand) galactic potential. In both cases, the  $5\times10^6\msun$ MO is initially at 8 kpc from the centre, with $f_{\rm circ}=0.3$, $\theta=i=15^\circ$, and initial velocity parallel to the disc plane. In the barred case, $\phi=24^\circ$ (the system has been rotated in the right-hand image, so that the coordinates of the initial MO position are the same in the barred and unbarred case). For each scenario, the three panels on the left-hand side show the projections of the orbit in time in three different directions, with $x-y$ being the plane of the disc. The initial 150 Myr of the orbital evolution are highlighted in red. The four panels on the right-hand side show, from top to bottom, (i) the distance of the MO from the centre of the system, (ii) the orbital energy per unit mass, measured in units of $4.301\times 10^4$ km$^2$ s$^{-2}$, and (iii - iv) the $x-y$ and $z$ components of the orbital angular momentum per unit mass, measured in internal units of 207.4 kpc km s$^{-1}$.  The dashed line in each plot marks the starting value for each of the displayed quantities. Interestingly, in the run with the bar, the MO gets dragged towards the centre faster thanks to the interactions with the bar, that allow it to reach the centre in less than a Hubble time, contrarily to the non-barred case.
    }
    \label{fig:orbit_example}
\end{figure*}


Fig.~\ref{fig:orbit_example} reports an illustrative example of how the bar may  affect the decay time-scale. An MO of  $5\times 10^6\msun$ on a relatively low-angular-momentum orbit, initially decaying from $r_0=8$  kpc with an initial inclination $i=15^\circ$, needs $<10$ Gyr to reach the centre if the bar is present, while it needs more than a Hubble time in the non-barred scenario. The plots also display some recurrent features of the orbital evolution: in the non-barred scenario, the evolution is way more smooth and predictable, contrarily to the stochastic evolution that characterizes the barred case; in both runs, the orbit circularizes and is dragged in the disc plane (as can be seen by looking at the different angular momentum components) at nearly kpc separation.

\subsection{Systematic orbital sampling}\label{sec:systematic_sampling}

\begin{figure*}
\centering
\includegraphics[ width=0.93\textwidth]{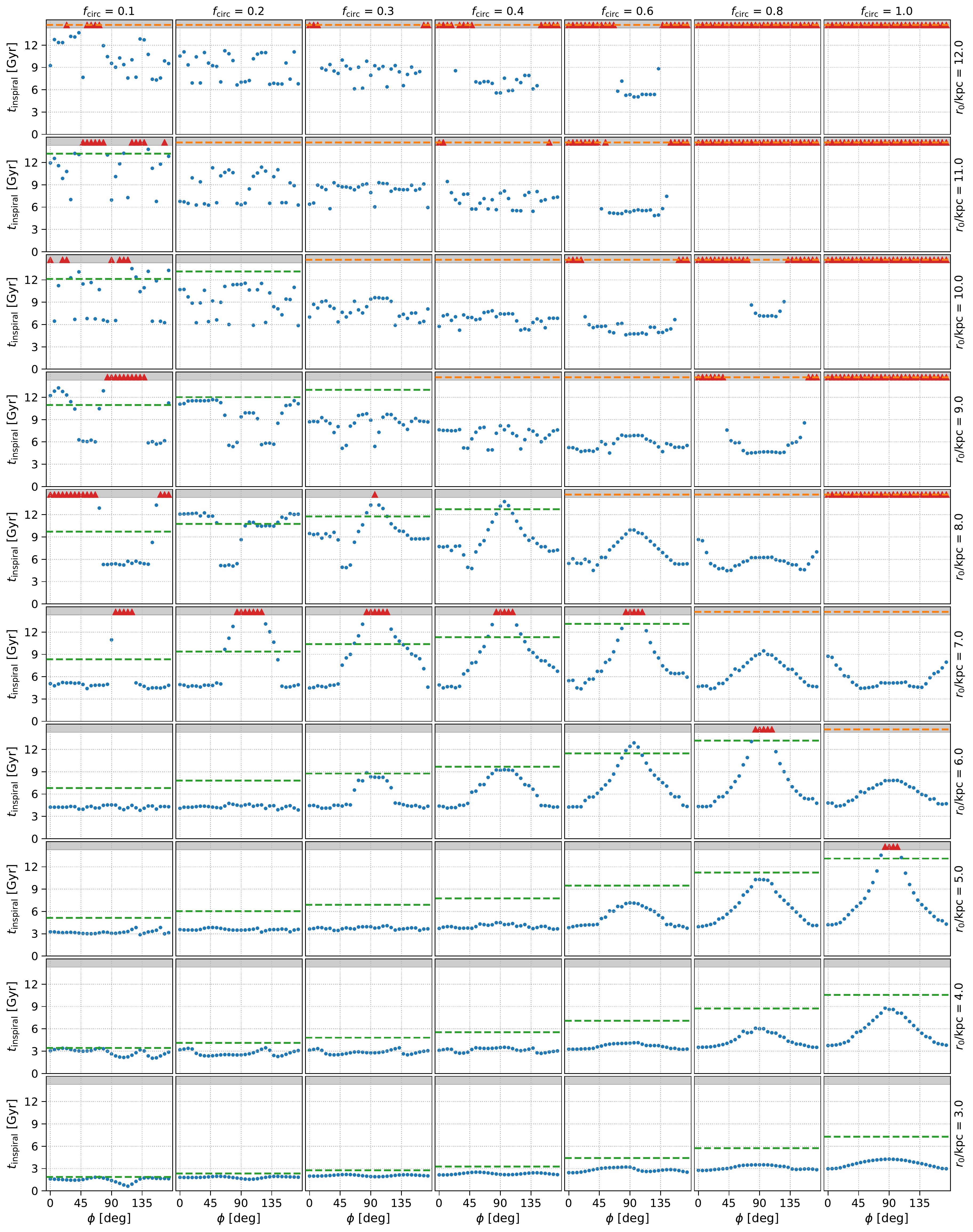}
    \caption{
    All runs shown here assume an MO coplanar with the disc and co-rotating with the bar and galactic disc (i.e. $\theta=\alpha=i=0$). The plots show the time for an MO of $5\times 10^6 \msun$  to reach the galaxy centre in a range of initial configurations: different rows consider a distinct initial separation from the centre ($r_0$ decreasing from 12 to 3 kpc, from top to bottom), whereas different columns imply an initial velocity expressed as a fraction of the circular velocity ($f_{\rm circ}$ increasing from 0.1 to 1, from left to right). In each panel, the green horizontal line marks the time needed by the MO to complete the inspiral in the non-barred galaxy; blue circles refer to the run with the bar and show the time needed for the MO to inspiral as a function of the phase $\phi$ (note that $\phi=0$ when the MO initially sits along the bar longest axis). The red triangles (and the orange dashed line, for the cases without a bar) mark the configurations for which the MO does not complete the inspiral within a Hubble time. 
    }
    \label{fig:td_inplane}
\end{figure*}

\begin{figure*}
\centering
\includegraphics[ width=0.97\textwidth]{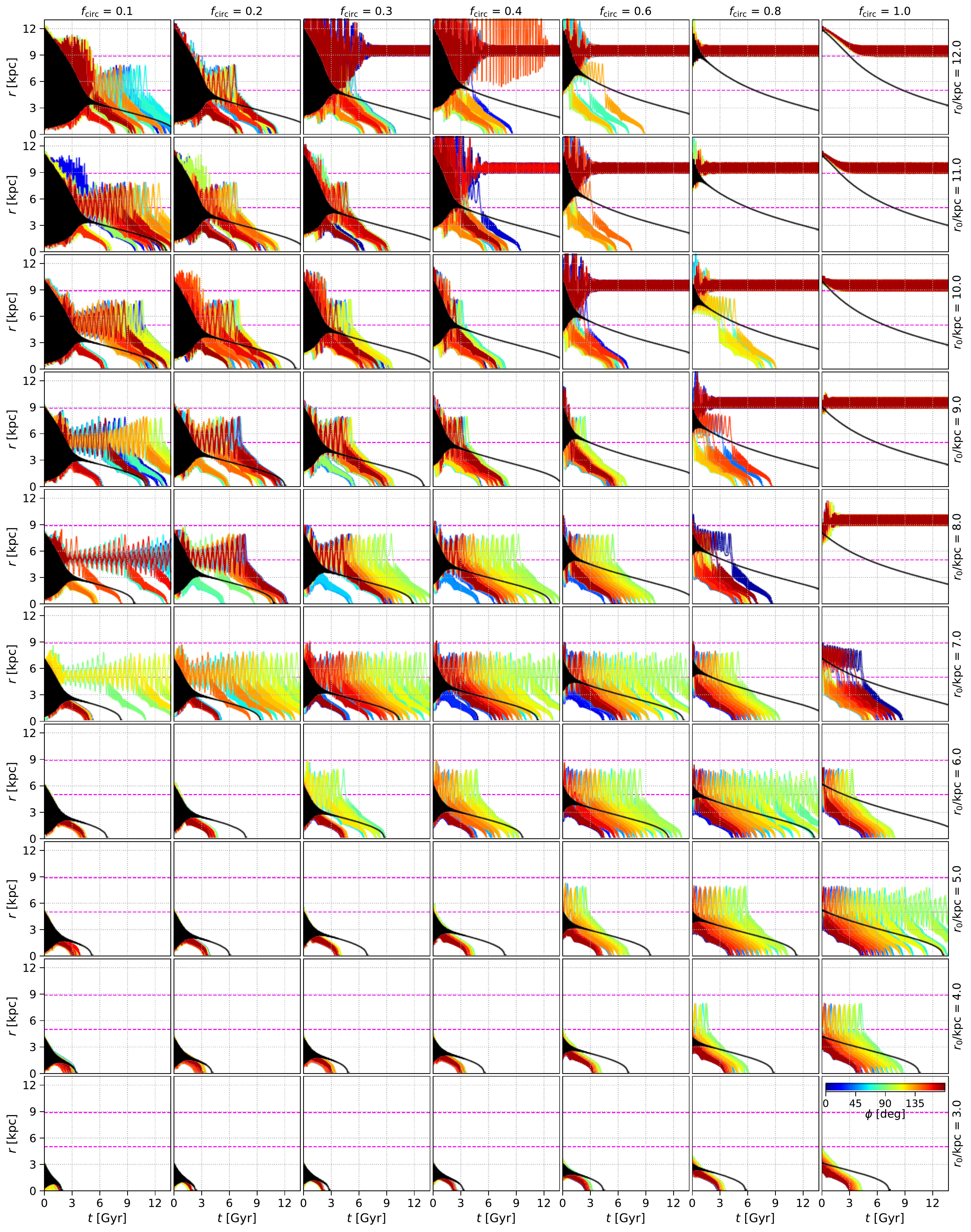}
    \caption{The matrix is the same as in Fig.~\ref{fig:td_inplane}. Each panel shows the distance of an MO from the centre as a function of time; the black line  refers to the run without the bar; the coloured lines in each panel refer to a run with the bar, each with a different phase, mapped in the colour-bar in the bottom-right panel. The two magenta horizontal lines mark the co-rotation radius (lower line) and the outer Lindblad resonance  radius (upper line). All runs shown here assume an MO coplanar with the disc and co-rotating with the bar and galactic disc.}
    \label{fig:mess0}
\end{figure*}

As a first test, we explore the orbital evolution of a $5\times 10^6\msun$ MO in the galaxy. This mass is a compromise between the typical mass an intruder MBH would have, if brought in the Milky Way by a minor merger, and the whole mass of the satellite galaxy that could host it. We find this   value to be a good compromise in order for a reasonable fraction of MO orbital decays to be completed in a Hubble time. 

\subsubsection{In-plane, prograde orbits}   

Fig.~\ref{fig:td_inplane}  shows the time needed by the MO to complete its inspiral for in-plane, prograde orbits (i.e. whose angular momentum has the same direction as that of the bar and the disc). 
In each sub-plot, the inspiral time is shown as a function of the phase $\phi$ (sampled as $\phi=0, 6, 12, ..., 174$ degrees) if the bar is present, while it is represented as a dashed line for the equivalent non-barred galaxy case. A more detailed view of the inspiral can be found in Fig.~\ref{fig:mess0},
which shows the MO  distance from the centre as a function of time for the same runs referenced in Fig.~\ref{fig:td_inplane}.  

The  effect of the bar on the orbital evolution and decay time-scale is particularly relevant for orbits that cross or initially remain close to the co-rotation radius, which roughly coincides with the bar major axis $a_{\rm bar}$ ($5$ kpc); at these scales, the bar reduces the decay time for orbits initialized near the edges of its major axis, while the decay time tends to be larger for initial phases near 90 degrees. 
As expected, the effect of the bar weakens for orbits which are initially close to the size of the second axis $b_{\rm bar}=2$ kpc, as can be seen by looking at the decay time-scales of MOs starting from small $r_0$ and small $f_{\rm circ}$ in Fig.~\ref{fig:td_inplane}. 
At scales of the order of the outer Lindblad resonance (Table~\ref{tab:resonances}), the interaction with the bar becomes less predictable and, in some cases, the bar keeps the MO out of $\approx 9$ kpc, preventing any inspiral and quashing the effect of DF, as can be  seen in Fig.~\ref{fig:mess0}.

\begin{figure*}
\centering
\includegraphics[ width=0.24\textwidth]{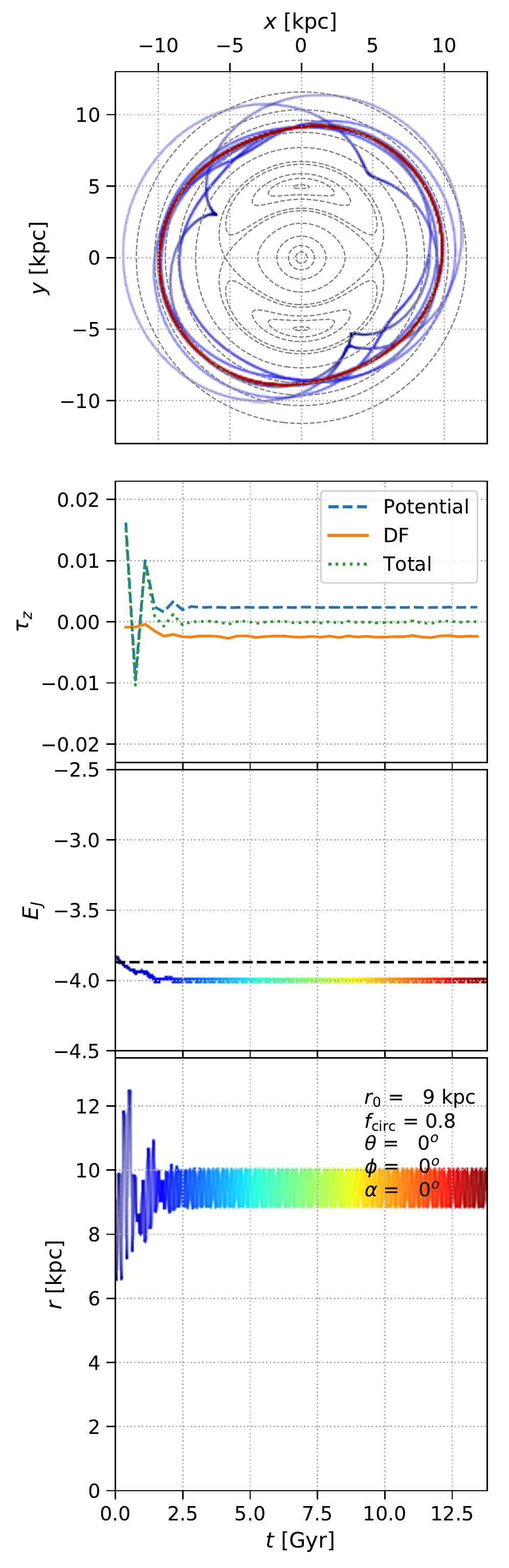}
\includegraphics[ width=0.24\textwidth]{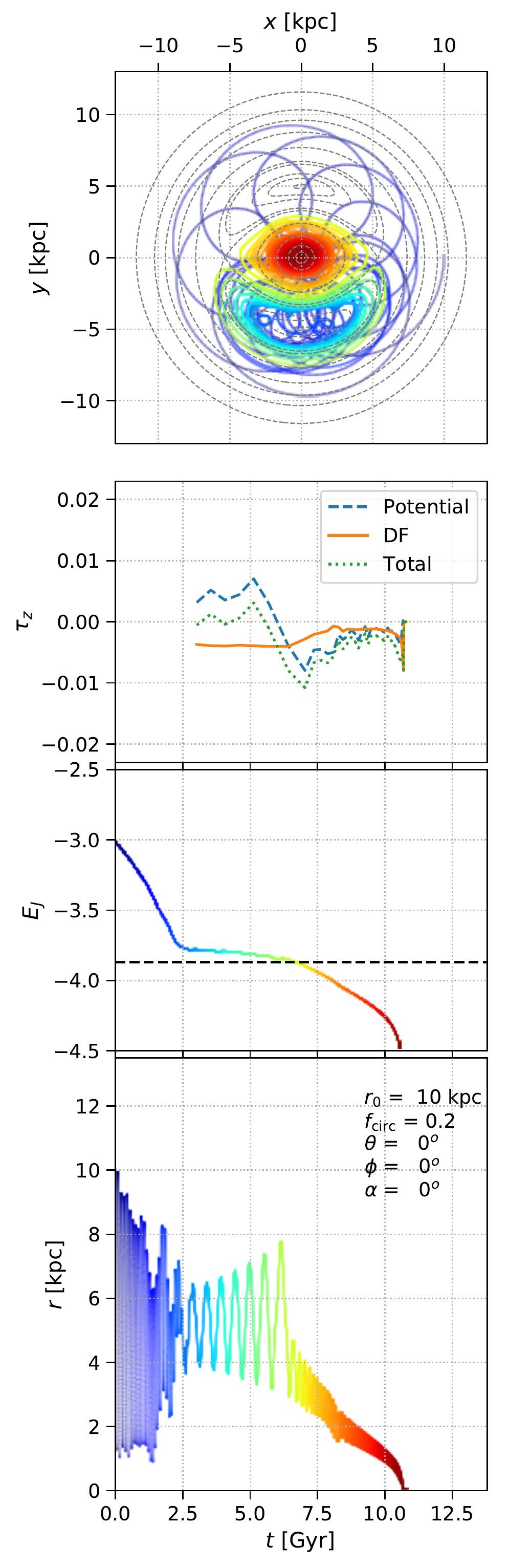}
\includegraphics[ width=0.24\textwidth]{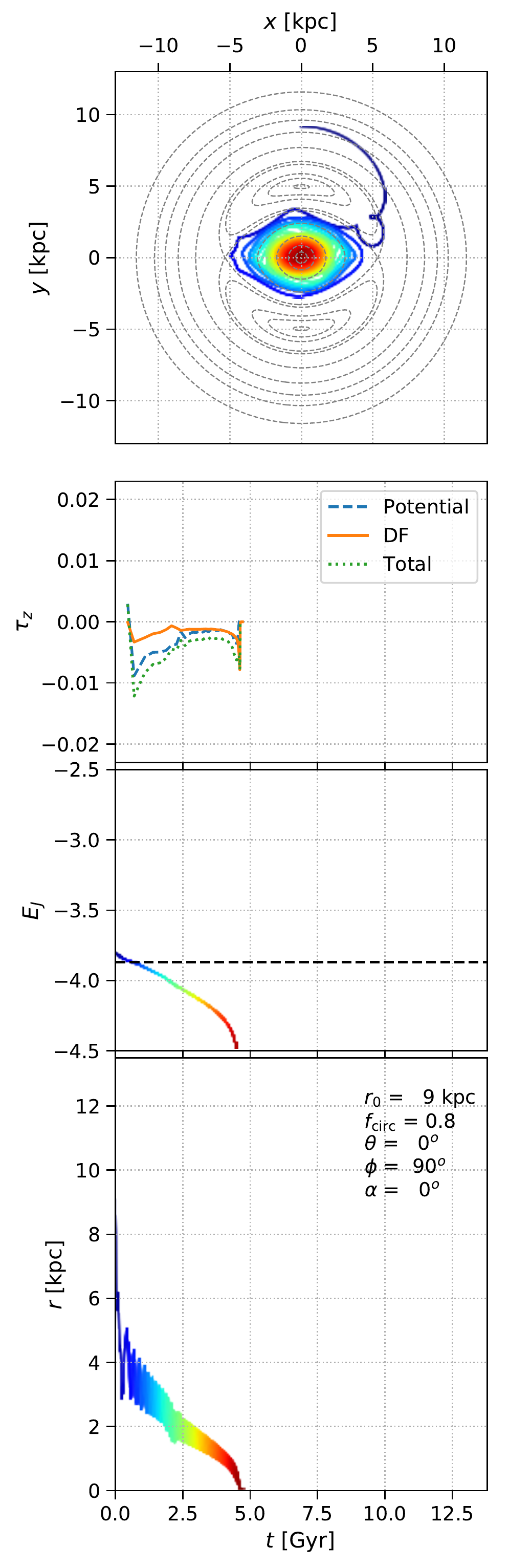}
\includegraphics[ width=0.24\textwidth]{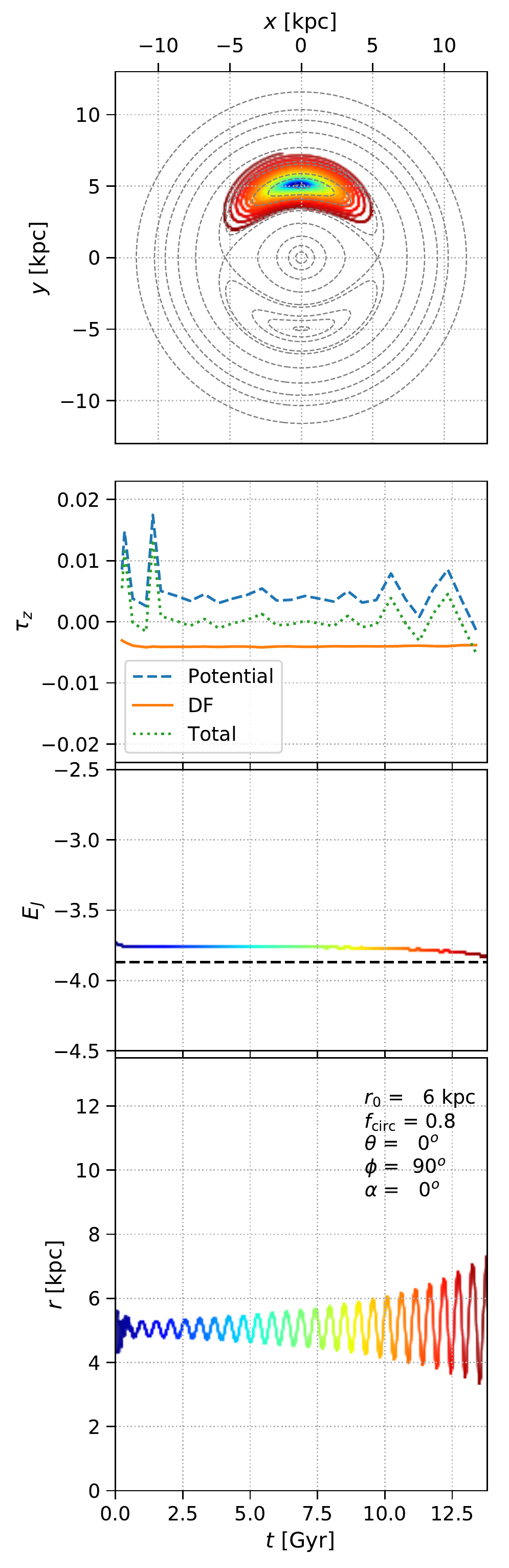}
    \caption{The plots show different aspects of the orbital evolution for a $5\times10^6\msun$ MO evolving in the barred potential. Each column refers to a different run, whose initialization variables are displayed in the bottom panels. The top plots show the MO orbital evolution in the rotating frame of the bar, and the colour associated to the line refers to a different time in the evolution, as mapped in the two bottom panels; the dotted grey lines are effective potential isocontours, the same as in Fig.~\ref{fig:bar_contour}. The second panel shows the $z$ component of the torque experienced by the MO averaged over a full azimuthal oscillation in the rotating frame, measured in  units of $4.4985\times 10^4$ kpc$^2$ Gyr$^{-2}$; we distinguish between the DF-induced torque, the global torque due to the `potential' of all components in the galaxy (see Footnote~\ref{fn:torque}), and the total torque experienced by the MO  (the sum of the aforementioned ones); note that, when the orbit is too irregular, it is almost impossible to get a proper orbit average of the torque, so this quantity is not shown for all time ranges. The third panel shows the value of the Jacobi integral (Eq.~\ref{eq:EJ}, measured in units of 4.301$\times10^4 $ km$^2$ s$^{-2}$) as a function of time, the black dashed horizontal line being the value of the effective potential at the saddle points. The bottom panel shows the distance of the MO from the galaxy centre as a function of time. All runs shown refer to prograde and in-plane MOs.
    }
    \label{fig:orbs_in_rotating_frame}
\end{figure*}

In order to better understand the aforementioned behaviour, we show in Fig.~\ref{fig:orbs_in_rotating_frame} 
the orbital evolution of the MO in the rotating frame for four different runs. The same Figure also shows the evolution of the different contributions to the $z$ component of the torque (averaged over a radial oscillation, $\tau_z$), of the Jacobi integral (Eq.~\ref{eq:EJ}), and the orbital radius of the MO. 
By examining Figs~\ref{fig:td_inplane}, \ref{fig:mess0}, and \ref{fig:orbs_in_rotating_frame},  we can see that the orbital evolution of MOs exhibits some recurrent behaviours: 
if an object starts from a large $r_0$, with $f_{\rm circ}\approx 1$, it may remain trapped in a nearly circular orbit near the outer Lindblad resonance, characterized by the same value for $E_{\rm J}\approx -4$ (in  units of $4.301\times10^4 $~km$^2$~s$^{-2}$), without experiencing any net decay. This behaviour is due to the positive bar-induced torque that, over a full orbit, counteracts the effect of DF, as  shown in  the first column of Fig.~\ref{fig:orbs_in_rotating_frame}.
Accordingly, Fig.~\ref{fig:mess0} clearly shows that several MOs starting from a large separation remain trapped there as they do not experience any net decay in about a Hubble time.

Fig.~\ref{fig:orbs_in_rotating_frame}  displays other typical configurations for the evolution: if the MO starts with an initial $E_{\rm J}$ larger than the maximum of the effective potential, then it can in principle explore the whole  galaxy. Owing to the drag of DF, though, $E_{\rm J}$ gets smaller and smaller, so that the orbit typically remains confined in a given region about one of the bar stable Lagrangian points (i.e. about one of the two effective potential maxima,\footnote{We stress that the maxima of the effective potential are not maxima of the gravitational potential, and therefore stable orbits can exist around these two Lagrangian points \citep[][]{Sellwood1993}.} or about the origin). This is, for instance, what is shown in the second column of Fig.~\ref{fig:orbs_in_rotating_frame}: the MO is initially wandering freely in the inner 10~kpc but, owing to the DF energy loss, it remains trapped about one maximum. While orbiting the maximum, it slowly decreases its $E_{\rm J}$ due to DF  and increases its eccentricity  (as it happens in the run in the fourth column in the same Figure; in that case, however, the MO spends a Hubble time orbiting the effective potential maximum), until it manages to go trough one of the two saddle points; from this moment, the DF-driven inspiral proceeds within the eye-shaped central crater, and the MO successfully inspirals towards the centre. Analogously, in the run shown in the third column of Fig.~\ref{fig:orbs_in_rotating_frame}, the MO wanders with an $E_{\rm J}$ close to the value of the effective potential at the saddle; since it immediately manages to pass close to one saddle point, its inspiral proceeds smoothly and effectively in the inner  eye-shaped hollow of the effective potential.

Note that the torque induced by DF and the global torque\footnote{\label{fn:torque} From this moment on, we will denote the torque experienced by the MO owing to the effect of the non-spherical and rotating galaxy potential (as opposed to the dissipative torque due to DF) as the \textit{global} torque. } may work against each other out of the central crater (as the rotating bar tends to increase the angular momentum of the MO), while they both promote the inspiral within the central hollow.

The aforementioned behaviours allow  for a better interpretation of the time-scales in Fig.~\ref{fig:td_inplane}: orbits with initial $\phi\approx0, 180^\circ$ starting their evolution with $r_0\approx5$~kpc and  $f_{\rm circ}\approx 1$ (or, analogously, with $r_0\gtrsim5$ kpc, but with $f_{\rm circ}<1$) start from a point that is very close to a saddle point, so that they can easily cross it and enter the region in which both $\tau_z$ from DF and the galaxy potential promote the inspiral. On the other hand, the evolution of these MOs, if it  starts from $\phi\approx90^\circ$, is necessarily delayed as they are initially `trapped'  near a potential maximum, and they remain there until their $E_{\rm J}$ becomes small enough so that they can  cross a saddle point and proceed with the central inspiral.

The behaviour of the decay time-scales for $r_0\gtrsim 8$ kpc is much less predictable, but it essentially boils down to understanding whether the MO starts oscillating about a potential maximum, so that it can eventually cross a saddle point and reach the centre, or whether it remains trapped in a circular orbit at the outer Lindblad resonance, not experiencing any net decay, as in the first column of Fig.~\ref{fig:orbs_in_rotating_frame}.
As a matter of fact, for orbits with $r_0\gtrsim7$ kpc and $f_{\rm circ}\gtrsim0.6$, the decay is typically possible if they start from $\phi\approx90^\circ$ 
(see, e.g. the case with $r_0=10$ kpc, $f_{\rm circ}=0.8$, or $r_0=12$ kpc, $f_{\rm circ}=0.6$), as  the effective potential at that location is slightly higher than that at the saddle point. On the other hand, the potential evaluated at the same initial radius for $\phi\approx0, 180^\circ$ is lower and therefore, for such values of $\phi$, the Jacobi integral is too small to allow for the crossing of the saddles. As a consequence, the associated orbits are more likely to remain trapped about the outer Lindblad resonance. %

Summarizing, the different behaviour of the MO orbiting near the co-rotation or outer Lindblad resonance  can be understood as follows. Near co-rotation, the MO would remain trapped about the ridges in the effective potential in absence of DF. As shown in the  right-most panels of Fig.~\ref{fig:orbs_in_rotating_frame},  the  orbit-averaged torque due to DF (which is always negative in this run) and  the oscillating  bar-induced one (which is positive, once orbit-averaged) nearly balance each other along the evolution; the two torques combine in such a way that, if the MO starts from near the top of the ridge, it then descends while  exhibiting wider and wider oscillations about the ridge top. As these oscillations grow larger, the non-averaged global torque grows in modulus due to the fact that the MO can get closer and closer to the bar. This  descent eventually brings the MO out of  the rim area so that it can cross the saddle point.

Orbits trapped about the outer Lindblad resonance  (see the left-most panels in Fig.~\ref{fig:orbs_in_rotating_frame}) behave quite differently. In there, both the bar torque and DF oscillate between positive and negative values along each orbit\footnote{DF can also induce an acceleration in rotating discs, see \citet{Chandrasekhar1942} and \citet{Bonetti2020}},  
and the net torque oscillates significantly as well. The net torque over each orbit is nearly zero, and the MO orbit does not drift along the evolution, once in the \textit{trap} orbit{, because of the angular momentum transfer from the bar to the MO that compensates for the loss due to DF}. Note that we tried to evolve the MO on this orbit for a hundred Hubble times, and we found no net decay. We further note that those trap orbits exist only for relatively light intruders, since DF becomes much stronger for significantly more massive MOs, overwhelming the bar-induced torque. 

\subsubsection{Counter-rotating orbits} \label{sec:counterrotating}

Figs~\ref{fig:td_inplane_counter} and~\ref{fig:mess180} of Appendix~\ref{sec:appendix_fig} show the analogous to Figs~\ref{fig:td_inplane} and ~\ref{fig:mess0}, respectively, but initializing the MO orbit so that it initially counter-rotates with respect to the galaxy angular momentum.
The bar impact on the decay time-scale is relatively modest in the counter-rotating cases, as the inspiral times remain very similar for runs with and without the bar. This is due to the following: when the MO is initially counter-rotating, its velocity relative to the bar is much larger than in the prograde case, so the bar does not manage to effectively torque the MO. This is true as long as the MO does not reverse the sign of its angular momentum. Indeed, a retrograde MO embedded in a rotating system has been shown to experience the so-called drag-towards circular co-rotation: this means that the MO would progressively lose angular momentum via DF, until its orbit gets very radial and its angular momentum reverses sign; from this moment on, DF would promote the circularization of the now prograde MO \citep{Dotti2007, Bonetti2020}. In the present framework, this means that the MO experiences little effect from the bar until its orbit turns to prograde: at that point the evolution can be assimilated to the prograde one, described above, and the effect of the bar becomes significant. 

We find that the MOs that switch the sign of their angular momentum earlier in the evolution are consistently found to take a longer time to complete their inspiral, for a given value of $r_0$ and $f_{\rm circ}$. This is likely due to the fact that, once the angular momentum reverses, circularization occurs promptly, thus the MO spends more time on a nearly circular orbit that does not reach the dense central regions where DF would be more efficient. On the other hand, if the angular momentum reversal never occurs or occurs when the inspiral is nearly completed, the MO orbit stays more eccentric, so that, along each orbit, it penetrates the denser regions near the centre, experiencing a stronger DF.
In addition, we found that the angular momentum sign reversal almost never happens if $f_{\rm circ}\gtrsim0.6$; this is likely due to the fact that the efficiency of DF is weaker if the relative velocity between the MO and the background is larger; given that retrograde, nearly circular orbits maximise this relative velocity, the effect of DF is weaker, thus circularization is not effectively promoted.

\subsubsection{Off-plane orbits}
Finally, we also explore the inspiral time-scale of  the same MO for off-plane orbits. We find that the bar effect   gets weaker as the initial orbit gets more off-plane. In general, the orbital evolution time-scale is very stochastic if the bar is present, and it is not easy to define a clear trend for the decay. We report the map that illustrates the decay time-scale in a set of off-planar runs in Fig.~\ref{fig:spectra000}. Note that  the off-plane MOs tend to get gradually dragged in the disc plane, where the evolution is analogous to what presented in the previous Sections.

\subsection{Monte Carlo orbital sampling}

\begin{figure*}
\centering
\includegraphics[ width=0.45\textwidth]{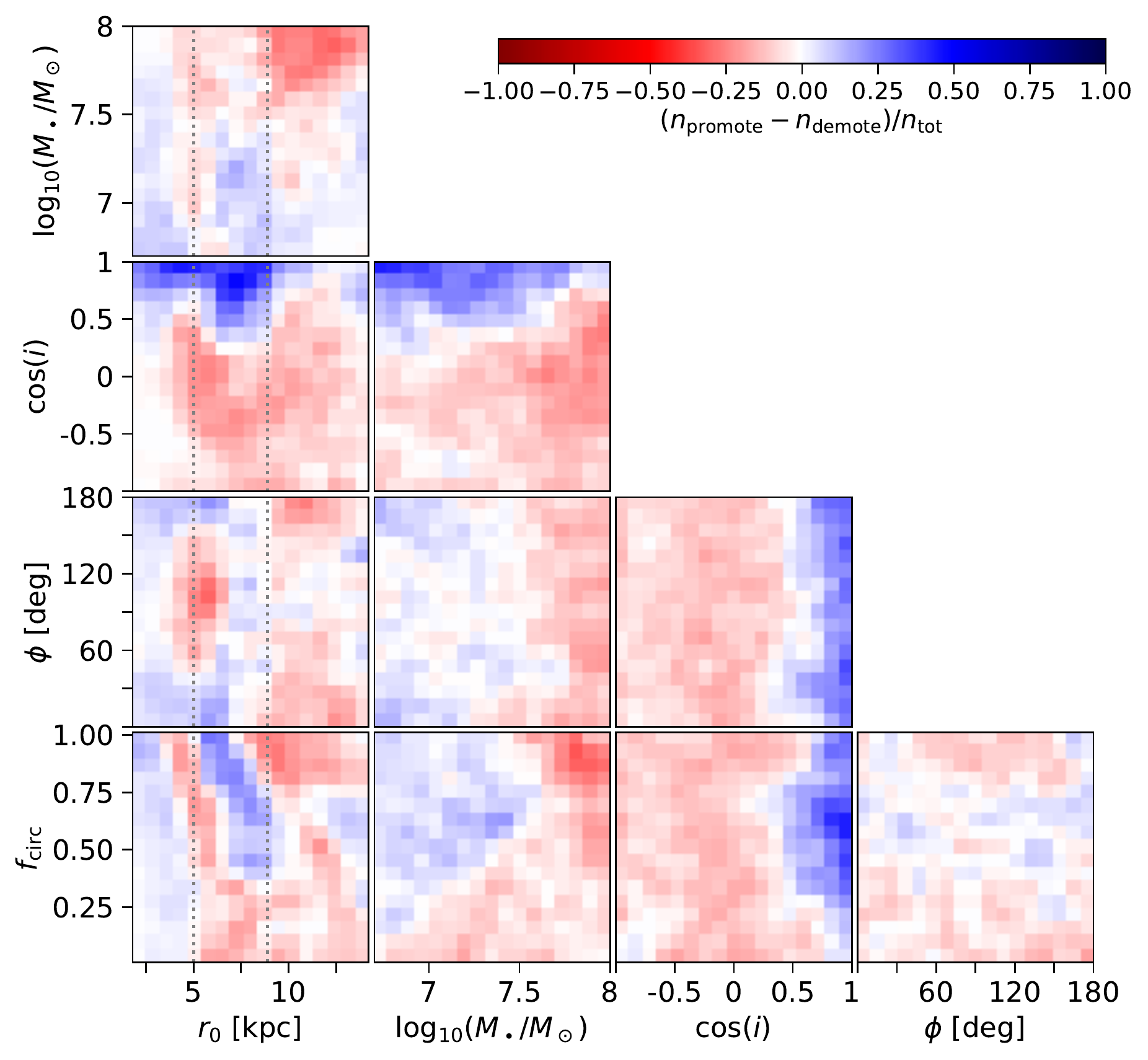}
\includegraphics[ width=0.45\textwidth]{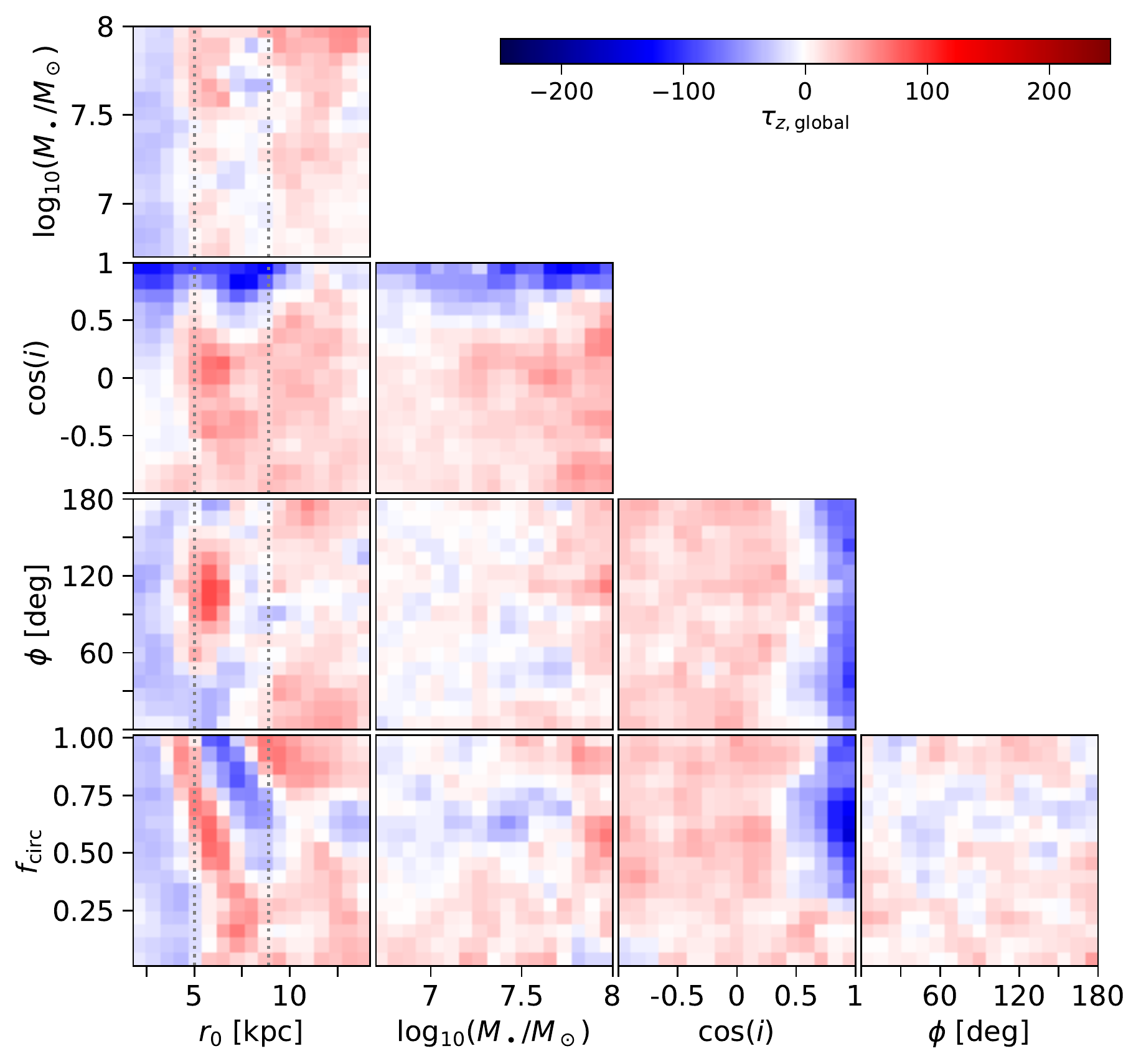}
\includegraphics[ width=0.45\textwidth]{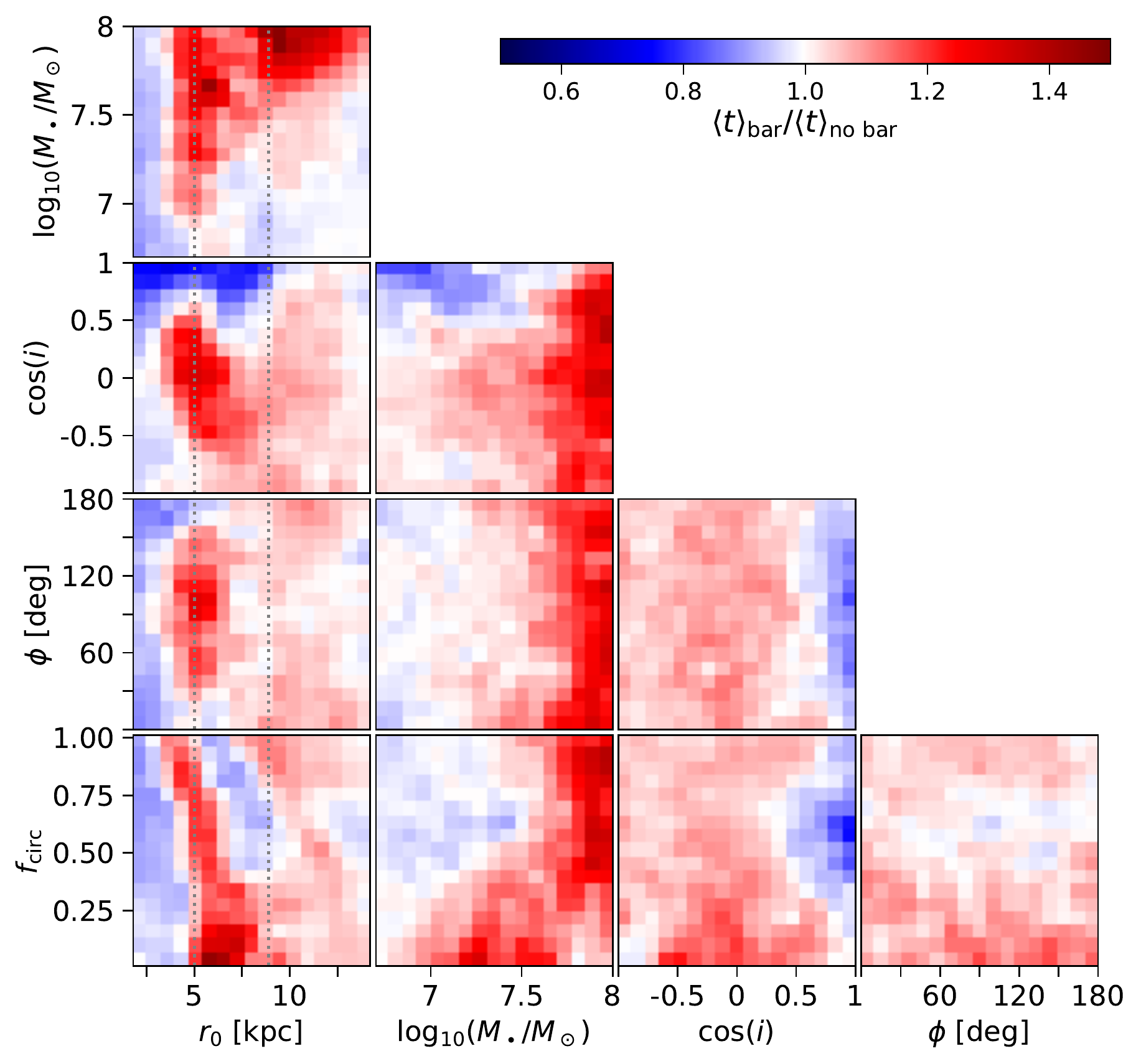}
\includegraphics[ width=0.45\textwidth]{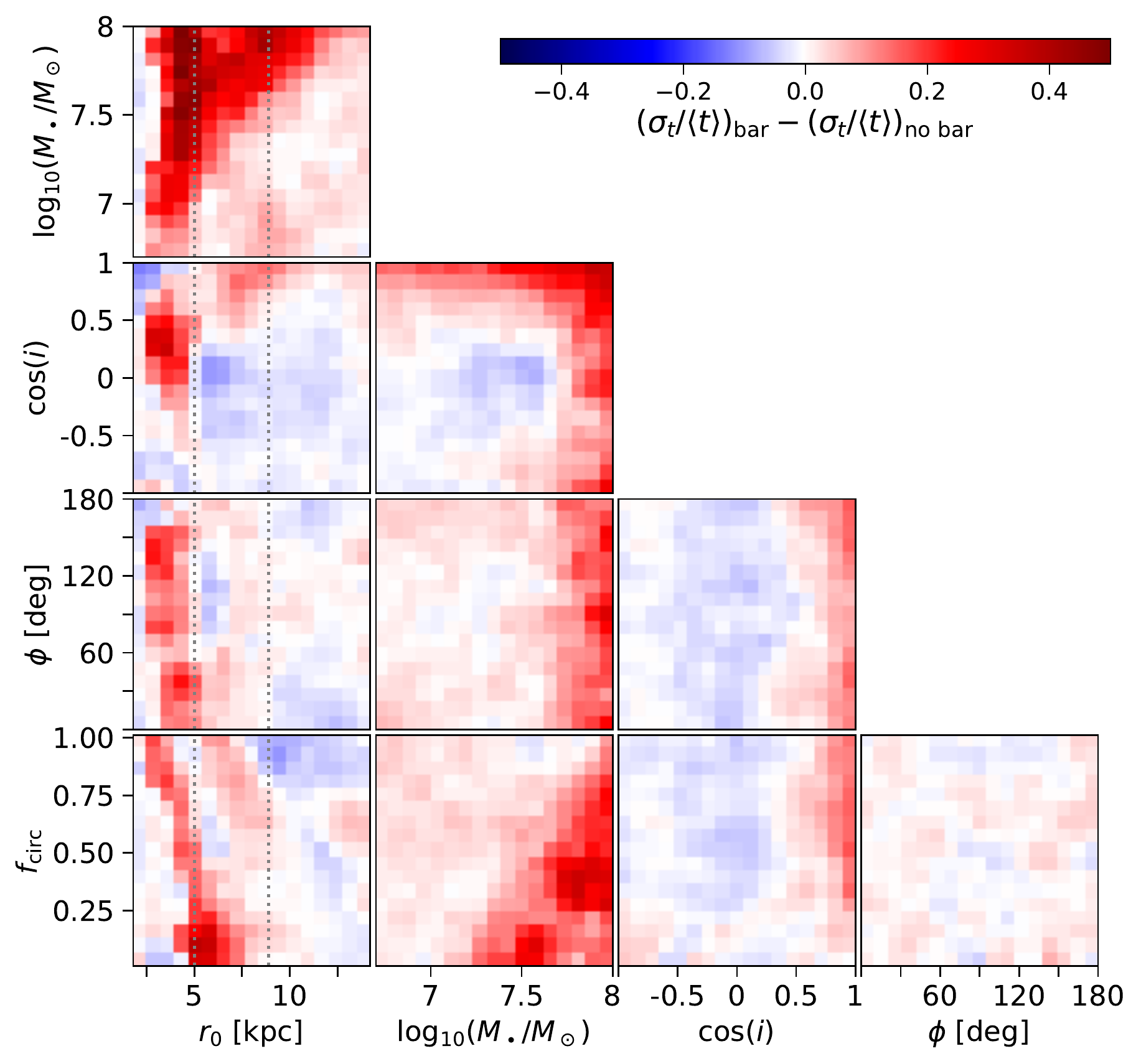}
    \caption{ The {\bf {top-left}} corner plot displays whether it is more probable that the bar promotes (blue) or demotes (red) the MO inspiral. More specifically, each region of the parameter space is colour-coded with the variable $f_{\rm b}=(n_{\rm promote}- n_{\rm demote})/n_{\rm tot}$, where $n_{\rm tot}$ is the total number of simulations in that given region of the parameter space, among which $n_{\rm promote}$ is the number of runs for which the barred inspiral time-scale is 0.75 or less times the non-barred inspiral; on the contrary, $n_{\rm demote}$ is the number of runs for which the \textit{non-barred} inspiral time-scale is 0.75 or less times the \textit{barred} inspiral. Runs taking more than a Hubble time are assumed to take infinitely positive time to inspiral; note that if  we assume runs taking more than a Hubble time to take exactly a Hubble time instead, the plot looks  similar.
    The {\bf {top-right}} corner plot shows the average magnitude of the $z$-component of the global torque in the barred runs: the time average of the torque is computed over each run, and  this value is averaged over all runs that belong to each different region of the displayed maps. 
    { The {\bf bottom-left} corner plot shows, for each given region of the parameter space, the ratio between the average MO inspiral time in the barred galaxy and the same quantity in the equivalent unbarred system, so that the red (blue) regions mark the portions of the parameter space in which the inspiral is slower with (without) accounting for the bar.
    The {\bf bottom-right} corner plot compares the degree of stochasticity associated with the ispiral time-scale in barred and unbarred systems, with the red (blue) colours showing the regions in which the inspiral time-scale gets more stochastic with (without) the bar.  More specifically, the colour map refers to the quantity $(\sigma_t/\langle t\rangle)_{\rm bar} - (\sigma_{t}/\langle t\rangle)_{\rm no\ bar}$, where $\langle t\rangle$, $\sigma_t$ respectively represent the average inspiral time and its standard deviation, and the subscript refers to whether we are considering runs with or without the bar. In the bottom panels, inspirals taking more than a Hubble time have been set to take 16 Gyr for the computation of the colour-coded quantities; we checked that this somehow arbitrary choice does not appreciably affect  our findings. 
    In all plots,} the vertical lines mark the co-rotation radius and the position of the outer Lindblad resonance. }
    \label{fig:competition}
\end{figure*}

In addition to the aforementioned simulations, we perform a series  of  runs initializing the MO so that its initial position is isotropic in a sphere of radius $r_0$, where $r_0$ is extracted uniformly in the range $[2, 14]$ kpc. We sampled the angle $\alpha$ uniformly between [0,360) degrees and $f_{\rm circ}$ uniformly between $[0.03, 1.0]$. We additionally sampled the MO mass in a log-uniform distribution between $5\times10^6\msun$ and $10^8\msun$, in order to understand the dependence of the inspiral also on the intruder's mass. For each extracted initial conditions,\footnote{Note that the distributions from which these quantities have been extracted for the Monte Carlo sampling have no claim to be  representative of a sample of MOs entering a real galaxy. } we run a simulation both in the barred and in the unbarred galaxy ($\approx 13,000$ runs).
Fig.~\ref{fig:competition}  shows {a set of corner plots: the left-hand ones show} whether the bar promotes (blue) or hinders (red) the inspiral for several combinations of parameters used in the orbit initialization. In particular, {in the top-left plot, } each area in the parameter space is colour-coded depending on the value of $f_{\rm b}=(n_{\rm promote}- n_{\rm demote})/n_{\rm tot}$, with $n_{\rm promote}$  the number of runs for which the barred inspiral time-scale is 0.75 or less times that of the non-barred case,  $n_{\rm demote}$  the number of runs for which the non-barred inspiral time-scale is 0.75 or less times that of the barred case (see the caption for more details), and $n_{\rm tot}$ the total number of runs in that given region of the parameter space. {The bottom-left panel, instead, is colour-coded with the ratio of the average inspiral time-scale in the barred and unbarred scenario. Both left-hand plots show very similar features.}
In general, there is a region near $r_0\approx5 $ kpc and $\phi\approx 90^\circ$ that shows the slow-down in the inspiral induced by the `trap'  near the effective potential maxima.
It is also clear that the bar tends to promote the inspiral of prograde MOs within the disc plane ($\cos(i)\approx1)$, at least within  the outer Lindblad resonance. Large MO masses are less likely to sink in the barred case, especially for large $r_0$  and $f_{\rm circ}$. Indeed,  the relative fraction  of inspirals that are not completed in a Hubble time with and without the bar within our complete Monte Carlo sample respectively amounts to 31 and 26 per cent; however, the same ratio amounts to 23.5 (9.6) per cent in the barred (non barred) scenario if we limit our analysis to MOs with $m_{\rm p}>10^{7.5} \msun$ and $r_0>7.5$ kpc.
This is likely due to the fact that the DF-induced deceleration increases linearly with the MO mass, whereas the effect of global torques is independent of the MO mass. More massive MOs thus sink more promptly in an axisymmetric, static potential where they experience DF alone; however, if the bar is present, the effect of DF is hampered  by global torques induced by the rotating triaxial structure: those typically hinder the inspiral at large scales.  \\

The results shown in the left-hand {panels} of Fig.~\ref{fig:competition} can be almost completely  explained in term of the $z$ component of the global (bar) torque for the barred cases. Indeed, in the {top-right} panel of the same Figure, we display the $z$ component of the (bar-induced) torque, time-averaged for every run, and then for all runs in a given region of the corner plot. This map nearly reproduces the left-hand ones, with averaged positive (negative) $z$ torques mapping the regions in which the bar promotes (demotes) the inspiral.

{Furthermore, the bottom-right panel of Fig.~\ref{fig:competition} compares the degree of stochasticity associated with barred and unbarred runs. In particular,  the plot is colour-coded according to the quantity  $(\sigma_t/\langle t\rangle)_{\rm bar} - (\sigma_{t}/\langle t\rangle)_{\rm no\ bar}$, where $\langle t\rangle$, $\sigma_t$ respectively represent the average decay time and its standard deviation within a given region of the parameter space, and the subscript refers to whether we are considering runs with or without the bar. This means that red (blue) regions mark the portion of the parameter space in which the decay time-scale is more stochastic with (without) the bar. In most cases,   the bar presence enhances the stochasticity in the same regions where the inspiral takes longer if the bar is present, and the bar average torque is positive. An exception is the region in which $\cos(i)=1$, mapping initially nearly prograde MOs. Those tend to have a faster inspiral in the barred case, at least for moderately light MOs starting from relatively small $r_0$; however, all coplanar runs accounting for the bar appear to have an enhanced degree of stochasticity, as for those runs the randomizing effect of the bar appears to be stronger.}

\section{Discussion and Conclusion}\label{sec:concl}

In this paper, we explored the orbital evolution of massive  objects (MOs)  in a barred Milky Way galaxy model, and we compared it to the evolution of MOs in an analogous, non-barred galaxy. We performed a large number of runs adopting a very accurate orbit integrator that features a careful treatment for the galaxy potential (including a bulge, a disc, a dark matter halo, and -- in some configurations -- a rotating bar) and careful treatment for dynamical friction (DF)  that properly recovers the results of $N$-body simulations even in rotationally supported galaxy discs \citep{Bonetti2020, Bonetti2021}.

We found that the presence of a typical galactic rotating bar, within an otherwise axisymmetric galaxy model, makes the MO orbital evolution more stochastic, and can significantly affect its orbital decay time-scale. In particular, the effect of the bar is more prominent for MOs that spend most of their evolution on a prograde orbit co-planar with the disc: in these situations, the inspiral time with and without bar can vary by a factor of a few.

These results are remarkable, especially considering that the chosen Milky Way-like galaxy did not feature an extremely prominent bar. Rather, its properties,
such as the mass, are compatible with the Milky Way bar including its pseudo-bulge component \citep{Portail2017}. The morphology of the considered system is analogous to that of many other spirals in the local Universe \citep{Kormendy2004, Drory2007},  in which pseudo-bulges are ubiquitous and are believed
to be originated from the bar itself, suggesting that our results should apply to typical late-type spirals, one of the most common class of galaxies
in the Universe.

In our runs, the bar presence often promotes the orbital decay but, in some  configurations (especially if the MO is initialized on a large prograde orbit co-planar with the disc), it does induce the stalling of an MO at large separations. This is in line with the results in \citet{Bortolas2020}: in their zoom-in cosmological simulation, MBHs were found to typically promptly inspiral when a bar develops in the host galaxy but, in one case, the bar instead scatters an MBH on a wide, large angular momentum orbit, hampering its further orbital decay.  \\

Our semi-analytical approach implies an idealized treatment of the host galaxy  and in particular of the DF drag, which is accounted for based on the \citeauthor{Chandrasekhar1943} implementation, 
a treatment that has its own limitations. Among those, the fact that the Coulomb logarithm entering the DF drag should be allowed to vary along the evolution  \citep{Petts2015}, rather than being kept fixed, was taken into account in our implementation; in addition, the standard DF treatment does not consider objects  moving faster than the MO in the braking effect, potentially resulting in 
an inaccurate evolution in some situations \citep[e.g.][]{Read2006, Antonini2012, Petts2015, Petts2016}. To constrain the impact of this approximation, we therefore 
checked that fast-moving stars do not crucially contribute to the friction in our implementation \citep{Bonetti2021}. It is also important to mention that  \citeauthor{Chandrasekhar1943}'s DF treatment assumes the response of the host galaxy to the passage of the MO  to be rather local, while in reality the whole host reacts to and \textit{resonates}  as a result of the MO perturbation \citep{Tremaine1984}. Taking into account this aspect is  very important \citep{Tamfal2020, Vasiliev2021} especially if the mass ratio between the host and the MO is not too far from unity. On the other hand, very low-mass MOs compared to the host, as those adopted in the presented study, are likely to result in a negligible global response from the host, so that the local treatment is  good enough \citep{Bonetti2021,Vasiliev2021}. Finally, our implementation of DF is relatively simplistic especially for the axisymmetric and triaxial structures.  In  particular, in the triaxial case, the bar may substantially affect  the main moments of the velocity distribution; as a result, the prescription adopted here may be systematically affected. Given that our results in the barred scenario are critically impacted by the interplay between DF and global torques, adopting a more accurate prescription for DF would affect the MO probability of approaching a resonance and remaining trapped into it or not. Nonetheless, it is clear that stochasticity plays a critical role in the orbital evolution of MOs in barred galaxies,  as demonstrated by the $N$-body simulations presented in Appendix~\ref{sec:appB}. Thus, the limitations of the presented DF implementation do not threaten the qualitative finding that bar resonances induce stochasticity in the orbital evolution of MOs.

Another important caveat concerns the temporal evolution of the galaxy (and its bar, when present): in our runs, the bar and galaxy properties were kept fixed, while in reality both would evolve significantly with time \citep[e.g.][]{Sellwood2014, Zana18a, Zana19}. A  \textit{live} bar, as opposed to the rigid bar potential adopted
here, may change its properties in time, possibly getting stronger 
(e.g. \citealt{Athanassoula2013}). This could not be taken into account in our semi-analytical treatment, and can only be addressed via devoted numerical simulations. Related to this, it is worth mentioning that the same galaxy merger that brings an MO in the outskirts of a  larger galaxy may influence the presence of a bar, possibly triggering its formation, delaying it or weakening/destroying a bar which is already in place \citep[e.g.][]{Pfenniger1990, Zana18b}.

Furthermore, (disc) galaxies may well feature further deviations from axisymmetry, the most obvious being spiral structures \citep[e.g.][]{Bertin1989}. The bar, when present, is generally the most prominent deviation, thus it would reasonably have the most relevant impact on the orbit of MOs; {spirals are generally  transient structures and may be strongly fluctuating, and a recent study shows their angular momentum transfer to the halo is negligible \citep{Sellwood2021}, suggesting the MO may be virtually unaffected by the spirals}. Additional torquing sources could be represented by clumps \citep[][]{Tamburello2017}, tidal perturbations to the galaxy \citep[][]{Bortolas2020} and many others; our simplified treatment represents a lower limit to the sources of stochasticity that may affect the inspiral of MOs. Finally, another important limitation of our study is the fact that we consider only point mass MOs with fixed mass and negligible extension, an approximation that is valid when the considered MO is an  MBH or a very compact cluster of stars.\footnote{As an example, a  nearly naked MBH might be wandering in the outskirts of a galaxy if e.g.  it was ejected from the centre as a result of the gravitational-wave recoil following the merger between two MBHs \citep[e.g.][]{Nasim2021}. Another possibility is to have a secondary galaxy that is severely ram-pressure stripped by the galaxy host, so that the intruder MBH remains nearly naked \citep{Capelo2015}. } If the MO were an extended and relatively diluted dwarf galaxy, or a stellar cluster, it would get trimmed by tidal forces along the evolution, depending on its properties with respect to the host's; modelling the effect of stripping however is beyond the scope of the present work.   
In spite of these limitations, our treatment allows to pinpoint the sole effect of the bar in the orbital evolution of an MO, and we defer the implementation of additional physical processes that may affect the inspiral to a forthcoming study.

To conclude, it is worth highlighting that we found the most massive MOs in our sample ($\gtrsim 10^{7.5}\msun$) that start their evolution from relatively large radii ($\gtrsim 8$ kpc) to be less likely to successfully complete the inspiral within the barred galaxy, compared to the axisymmetric case; when the bar is present, we find that the number of stalled MOs may double. This aspect is particularly relevant considering that, in a realistic scenario, one expects a relatively massive intruder galaxy to start interacting from large separations. 
In particular, MOs might be delivered by minor mergers, in the form of cores of
a galaxy companion, and since these might be easily dropped at the outskirts of the galaxy when the host companion is  tidally dissolved (e.g. \citealt{Callegari2009, Callegari2011}), this outcome might not be rare. {Our runs suggest that, in the limit of minor mergers that was probed in this work, the most massive MOs which are the most affected by DF are also those whose large-scale inspiral is most effectively hindered by the bar, implying that barred galaxies involved in minor mergers are likely to feature lower  rates of MBH mergers. However, the  statistical relevance of this } should be evaluated with the help of cosmological simulations or
semi-analytical models of galaxy formation modeling a large sample of systems. {In general, we stress that}
the presence of bars and other deviations from axisymmetry should  be taken into account when exploring the accretion of galaxy satellites onto more massive systems, and when making predictions of MBH mergers, as the rates of gravitational wave driven MBH coalescences are closely connected with the efficiency of inspiral of their parent systems. For example, MBHs detectable by LISA should be abundant in the mass range $10^5-10^7 M_{\odot}$, which is the typical
mass of MBHs hosted by late-type spirals, the class of galaxies in which the dynamical processes discussed in this paper is most relevant.  In this context, our semi-analytical framework can be implemented into semi-analytical models of galaxy and MBH formation and evolution, to better evaluate the impact of bars on the formation and evolution time-scales of MBH binaries, which is critical in estimating the formation rate of gravitational-wave sources detectable by forthcoming low-frequency gravitational wave facilities such as LISA \citep{Bonetti2019}.

\section*{Acknowledgements}

We thank the anonymous referee for their useful comments and suggestions.
We warmly thank Eugene Vasiliev for his help and support for the usage of the AGAMA tool. EB and AS acknowledge support from the European Research Council (ERC) under the European Union's Horizon 2020 research and innovation program ERC-2018-COG
under grant agreement N.~818691 (B~Massive). EB, PRC, and LM 
acknowledge support from the
Swiss National Science Foundation under the Grant 200020\_178949.
MD, MB, and AL acknowledge funding from MIUR under the grant PRIN 2017-MB8AEZ.

\section*{Data Availability Statement}
The data underlying this article will be shared on reasonable request to the corresponding author.




\bibliography{bibliography} 


\appendix

\section{Decay time-scales for inclined or counter-rotating runs}\label{sec:appendix_fig}

\begin{figure*}
\centering
\includegraphics[ width=0.95\textwidth]{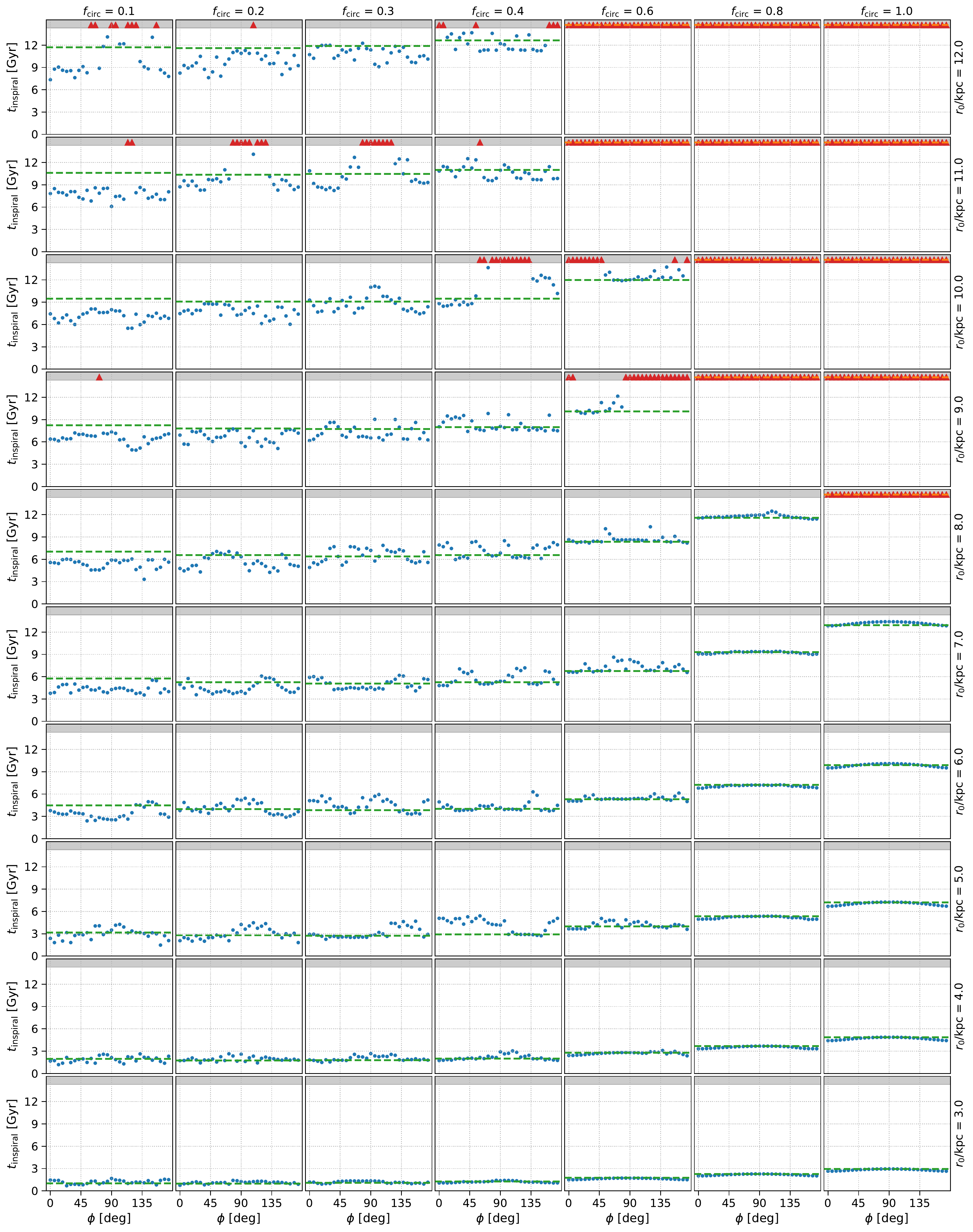}
    \caption{Same as Fig.~\ref{fig:td_inplane} but for MOs initially counter-rotating with respect to the bar and disc angular momentum. Counter-rotating orbits seem to be much less affected by the presence of the bar, except for very radial orbits starting from large separations, which have time to experience the drag towards circular co-rotation and later evolve coplanarly, so that their late evolution can be assimilated to the co-rotating co-planar cases. }
    \label{fig:td_inplane_counter}
\end{figure*}

\begin{figure*}
\centering
\includegraphics[ width=\textwidth]{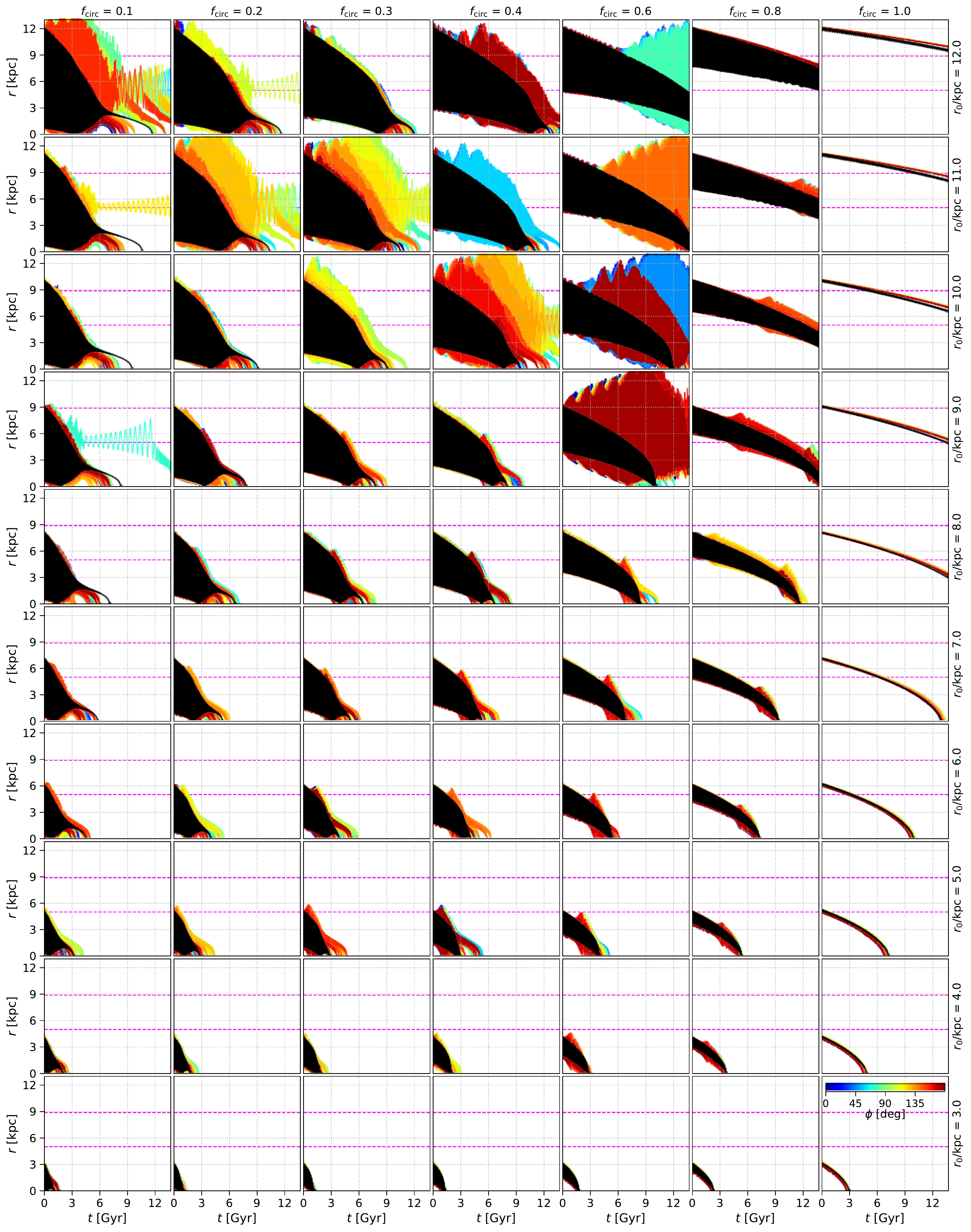}
    \caption{Same as Fig.~\ref{fig:mess0} but for MOs initially counter-rotating with respect to the bar and disc angular momentum. }
    \label{fig:mess180}
\end{figure*}

\begin{figure*}
\centering
\includegraphics[ width=0.91\textwidth]{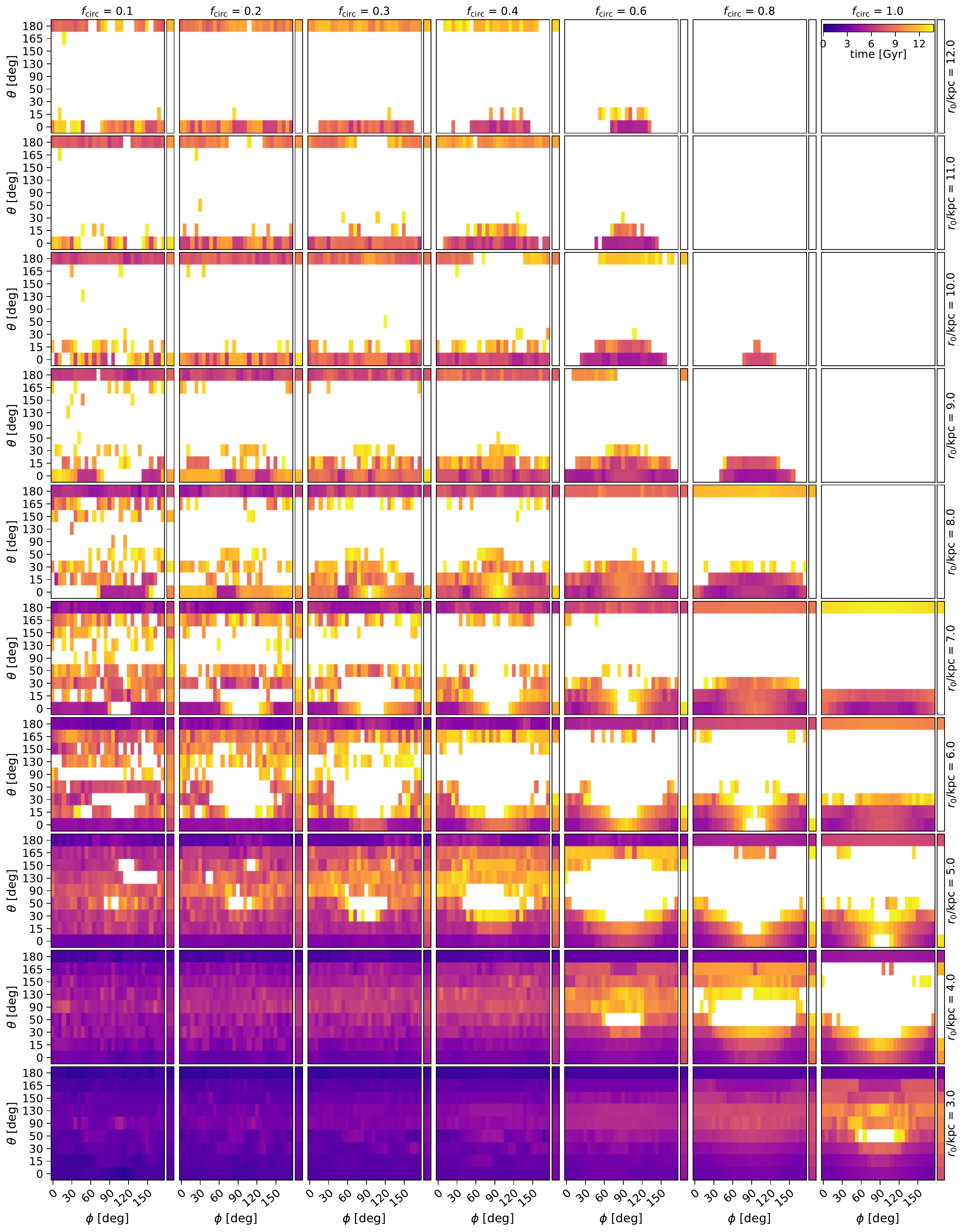}
    \caption{In each plot, the colour refers to the time needed for the MO to complete the inspiral, from blue (0~Gyr) to yellow (13.7~Gyr, the colour-bar is shown in the top right plot); the white regions refer to runs in which an MO does not complete the inspiral within a Hubble time.
    Each column of subplots corresponds to a different $f_{\rm circ}$ (growing from left to right, i.e. the initial orbit is more and more circular from left to right); each row corresponds to a different initial MO separation,  $r_0$, which grows from bottom to top.
    We show the decay time as a function of the initial phase $\phi$  and the angle $\theta$; note that the angle $\theta$ is sampled non uniformly to have a finer grid near the disc plane. The narrow coloured strip next to each sub-plot represents the reference colour-coded inspiral time for the runs without the bar (which have no phase dependence). All plots refer to the case for which $\alpha=0$ (meaning that the initial MO velocity is always parallel to the disc plane), and assume an MO of $5\times 10^6\msun$. 
    }
    \label{fig:spectra000}
\end{figure*}

In this appendix, we collect further figures that illustrate the decay time-scale and orbital evolution for MOs of $5\times10^6\msun$ for a larger region of the parameters space. Fig.~\ref{fig:td_inplane_counter} shows the decay time-scale for counter-rotating, in-plane  orbits sampled as in Fig.~\ref{fig:td_inplane}; Fig.~\ref{fig:mess180} shows the associated radial separation as a function of time.
Fig.~\ref{fig:spectra000} maps  the inspiral time-scale of the same MO  for different, sampled values of $r_0$, $\theta$, and $\phi$. That is, the MO is now allowed to orbit out of the disc plane.
The results associated to the aforementioned plots are briefly presented in Sec.~\ref{sec:systematic_sampling}. 

\begin{figure}
\centering
\includegraphics[ width=0.4\textwidth]{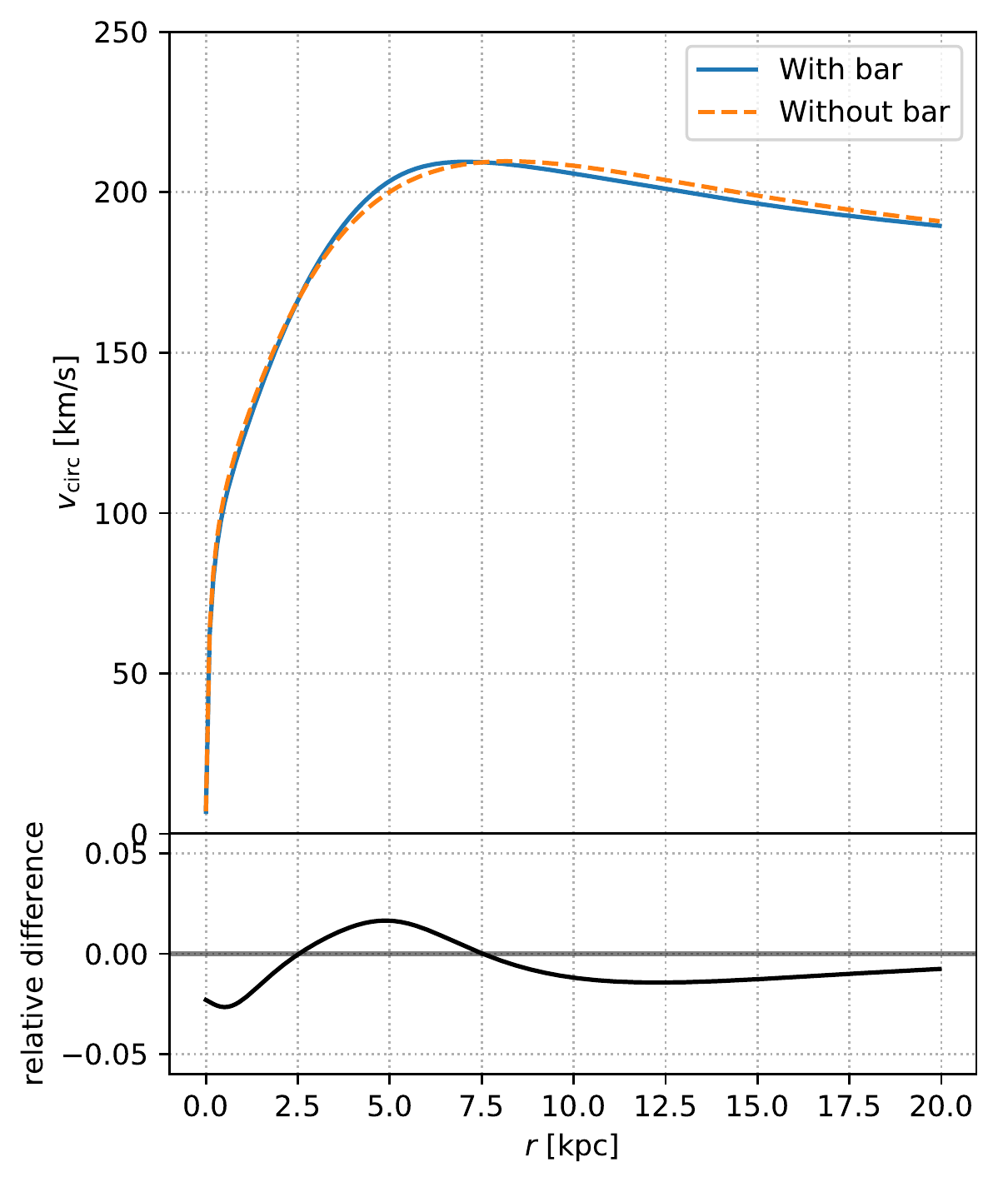}
    \caption{Comparison between the in-plane rotation curve for the model accounting for the bar (blue solid line in the top panel) and for the one in which the bar is not in place, thus its mass is re-distributed between the bulge and disc component (orange dashed line). The relative difference between the two rotation curves is shown in the bottom panel and its absolute value is always below 4 per cent. The relative difference is computed as $(v_{\rm circ,bar}-v_{\rm circ,no\ bar})/v_{\rm circ,bar}$, with $v_{\rm circ,bar}$ being the circular velocity associated with the barred potential (note that here the velocity is obtained by averaging over all possible bar orientations) and $v_{\rm circ,no\ bar}$ being the circular velocity in the galaxy not featuring a bar.}
    \label{fig:rot_curve}
\end{figure}

Finally, Fig.~\ref{fig:rot_curve} compares the rotation curve of the barred and unbarred galaxy models adopted in the text. The two quantities deviate by up to 4 per cent.

\section{Comparison with numerical simulations}\label{sec:appB}

\begin{figure*}
\centering
\includegraphics[ width=0.8\textwidth]{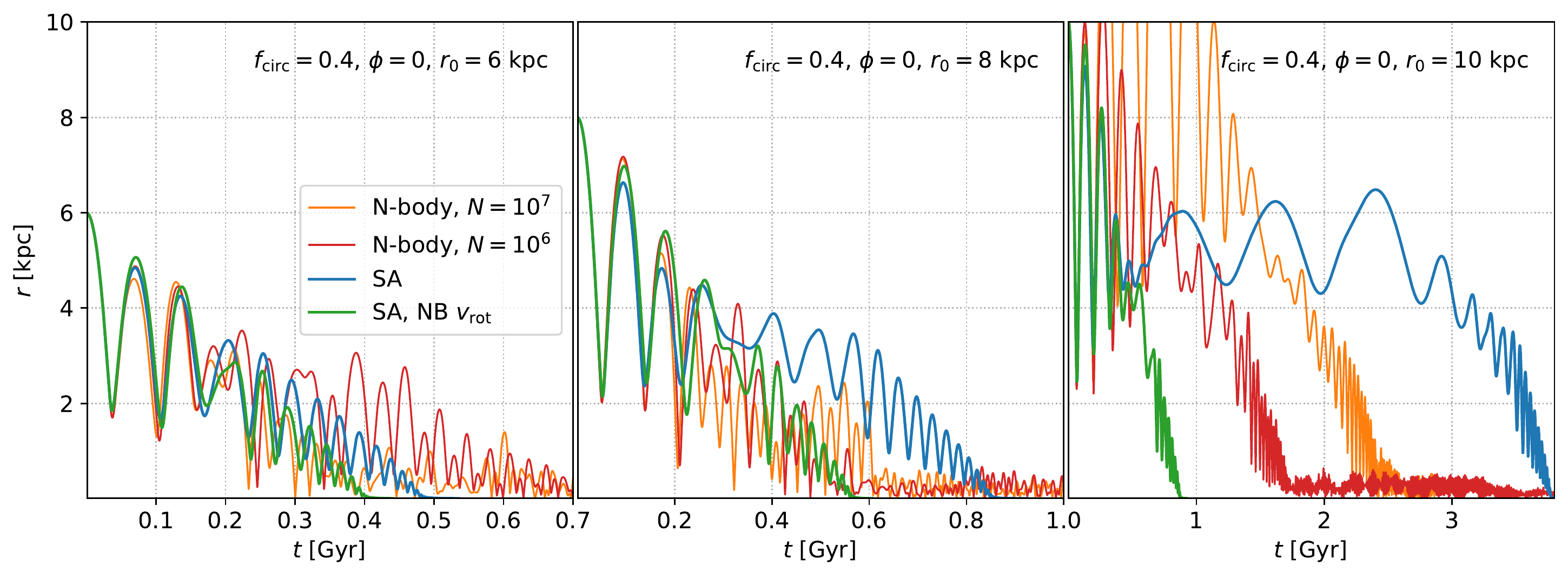}
    \caption{The plot displays the distance as a function of time of the $5\times10^7\msun$ MO for the two $N$-body simulations with a different resolution ($N=10^7$ and $10^6$ particles for the orange and red line, respectively) and for the semi-analytical runs employing the rotational velocity obtained either via the standard method adopted in the remaining of the paper (blue line) or from the initial conditions of the $N$-body run (green line). The initializing parameters for the orbit are shown at the top of each panel.}
    \label{fig:comp_nbody}
\end{figure*}

In order to test the validity of our implementation in the orbital integration, we performed a set of $N$-body runs and compared them with the results obtained via the semi-analytical orbital integrator adopted throughout the paper. The $N$-body simulations have been performed using \textsc{gizmo} \citep{gizmo}. The initial conditions for the run were obtained by means of the \textsc{agama} package \citep{Vasiliev2018} and feature a disc, bulge, and dark matter halo whose properties are the same displayed in Tab.~\ref{tab:mwstruct}, except for the disc mass which was increased to $3.3\times10^{10}~\msun$ to enhance the stability of the model. Here the bar was implemented as a Ferrers  potential \citep{Pfenniger1984}, in place of the softened needle adopted in the rest of the present study; in fact, the Ferrers potential (contrarily to the softened needle one) was available in the \textsc{agama} toolkit and allowed us to generate equilibrium initial conditions. The bar properties were chosen so that the total mass is $1.5\times10^{10}\msun$, its axes ratios are $(a,b,c)_{\rm bar} = (5,2,1)$~kpc and its initial rotational frequency is 40 km s$^{-1}$ kpc$^{-1}$; these choices allowed to obtain a more stable system. We performed two different simulations, with the same properties but with different resolution, i.e. one with  $N=10^6$ particles and  10 pc of spatial resolution, and another one with  $N=10^7$ and 3 pc of spatial resolution. In both situations, the dark matter halo was included as a rigid analytical potential, while the bar, disc, and bulge structure were  modelled with live particles. We put a total of 16  MOs  of $5\times 10^7\msun$ (i.e. respectively $\approx 10^3$ and $10^4$ times more massive than the other particles in the run) in the run, whose orbits were initialized as follows: $r_0=\{4, 6, 8, 10\}$ kpc, $f_{\rm circ} = \{0.4,0.8\}$, and $\phi=\{0, 90\}$, for a total of 16 configurations. In addition, we switched off the gravitational interaction between these MOs, so that each of them could be considered to be evolving independently from the others, allowing us to obtain a good statistical sample at a relatively limited computational cost.\footnote{A similar strategy was proposed and adopted for the first time by \citet{Bortolas2020}.} We simulated the evolution of the same MOs, within the same underlying potential (the Softened Needle potential for the bar was here replaced by the Ferrers potential), via the semi-analytical orbital integrator, switching off the dynamical friction from the halo for consistency with the $N$-body run. The semi-analytical orbit integration was performed adopting two different choices for the rotational velocity $\mathbf{v}_{\rm rot}(R)$ entering via $\mathbf{v}_{\rm rel}$ in  Eq.~\ref{eq:adf_disc}; in one case, we computed it as in \citet{Bonetti2021} (i.e. as it was done for all the other semi-analytical integrations in the main body of the paper); in addition, we also computed  $\mathbf{v}_{\rm rot}(R)$ by extracting it from the initial conditions of the simulation and averaging it in the azimuthal direction considering particles within a thin slice (of total height 60~pc) about the disc plane.

The MOs distance in time for a sub-sample of our  initial conditions and for each of our integration methods is shown in Fig.~\ref{fig:comp_nbody}. The figure shows that stochasticity overall plays a central role in the evolution of MOs. In fact, even changing the resolution of the simulation, the inspiral time-scale and overall orbital decay change significantly. This implies that we cannot expect a one-to-one comparison between our semi-analytical orbital integrator and the $N$-body simulation. Our results also clearly suggest that the orbits are very chaotic, and just a small perturbation along the evolution results in a different dynamics and orbital decay. This also implies that, although our semi-analytical prescription for DF may not capture completely the physics underlying this phenomenon, uncertainties associated with its modelling are most likely a secondary effect, with stochasticity being the main actor in determining the evolution. This strongly supports the fact that bars may induce very erratic dynamics in the orbital decay of MOs.

\bsp	
\label{lastpage}
\end{document}